\documentclass[9pt,twocolumn,twoside]{osajnl}
\usepackage{tikz,wrapfig}
\usetikzlibrary{decorations.pathmorphing,patterns}
\usepackage{siunitx}
\usepackage{booktabs} 
\usepackage{float}
\usepackage{mathtools,amssymb}
\usepackage{bookmark}
\usepackage{subcaption}

\journal{ao} % Choose journal (ao, aop, josaa, josab, ol, pr)

\setboolean{shortarticle}{false}
\title{Influence functions for a hysteretic deformable  mirror  with  a  high  density  2D  array  of  actuators}

\author[1,*]{A.E.M. Schmerbauch}
\author[1]{M.A. Vasquez-Beltran}
\author[2]{A.I. Vakis}
\author[3]{R. Huisman}
\author[1]{B.~Jayawardhana}

\affil[1]{Engineering and Technology Institute Groningen - Discrete Technology \& Production Automation, Faculty of Science and Engineering, University of Groningen, 9747AG Groningen, The Netherlands}
\affil[2]{Engineering and Technology Institute Groningen - Computational Mechanical and Materials Engineering, Faculty of Science and Engineering, University of Groningen, 9747AG Groningen, The Netherlands}
\affil[3]{Netherlands Institute for Space Research, Groningen, The Netherlands}

\affil[*]{Corresponding author: a.e.m.schmerbauch@rug.nl}

\dates{Appl. Opt. 59, 8077-8088 (2020)}

 \doi{\url{https://doi.org/10.1364/AO.397472}}

\begin{abstract}
We present modeling and analysis of a hysteretic deformable mirror where the facesheet interacts with a continuous layer of piezoelectric material that can be actuated distributively by a matrix of electrodes through multiplexing. Moreover, a method for calculating the actuator influence functions is described considering the particular arrangement of electrodes. The results are presented in a semi-analytical model to describe the facesheet's deformation caused by a high density array of actuators, and validated in a simulation. The proposed modeling of an interconnection layout of electrodes is used to determine the optimal pressures the actuators have to exert for achieving a desired surface deformation.
\end{abstract}

\setboolean{displaycopyright}{true}

\begin{document}

\maketitle

\section{Introduction}
Deformable mirrors (DMs) are instruments used for the correction of light wavefront aberrations in many imaging and nonimaging applications such as three-dimensional (3D) imaging to increase the realism of depth perception \cite{Zawadzki2005,Shain2017}, microscopes to correct static lenses \cite{Booth2007}, medical applications \cite{Fernandez2001}, or industrial applications like laser material processes \cite{Vdovin2001}. In general, DMs are distinguished in segmented and continuous facesheet mirrors, and can be further classified by means of their actuator type that is mounted below the reflective top layer to deform the mirror surface. Depending on the application, various actuator technologies are used which include, for example, piezoelectric \cite{Hardy1998, Sato1980, Wlodarczyk2014}, electrostatic \cite{Cornelissen2010}, thermal \cite{Vdovin2002,Huang2015}, magneto-restrictive and shape memory alloy actuators \cite{Gilbertson1996} as well as voice coil/ reluctance actuators \cite{Biasi2010,Madec2012}. 
Furthermore, DMs are applied in adaptive optical systems and key instruments for space telescopes. When a distorted incoming wavefront arrives at the telescope, a wavefront sensor is used to measure the wavefront distortion and subsequently used to adjust the shape of DM in order to correct the distorted wavefront. Future large space telescopes like LUVOIR \cite{team2019luvoir} use coronographic instruments for high-contrast imaging of exoplanets. Although thousands of exoplanets have been identified, the current state of technology limits our capability in measuring and understanding these exoplanets beyond their mass, radius, orbital period and distance to the host star. To overcome these challenges and provide the required capabilities for a direct exoplanet imaging space mission, DMs strive among others after high actuator density, meaning that the number of actuators must be increased to the maximum that can still guarantee practical operability for wire bonding, harness and electronics. DMs usually have a number of actuators ranging from 100 to 6000 but rarely higher \cite{Madec2012, Freeman1981, Riaud2012}. One of the major limitations for employing the mirrors with a large number of actuators on a space mission is the reliability of the associated cable harness and electronics. If every actuator has to be driven continuously to hold a specific position, a dedicated channel consisting of a digital-to-analog converter and a high voltage amplifier is required per actuator, resulting in bulky electronics. 

\begin{figure}[htbp]
    \centering
    \includegraphics[width=\columnwidth]{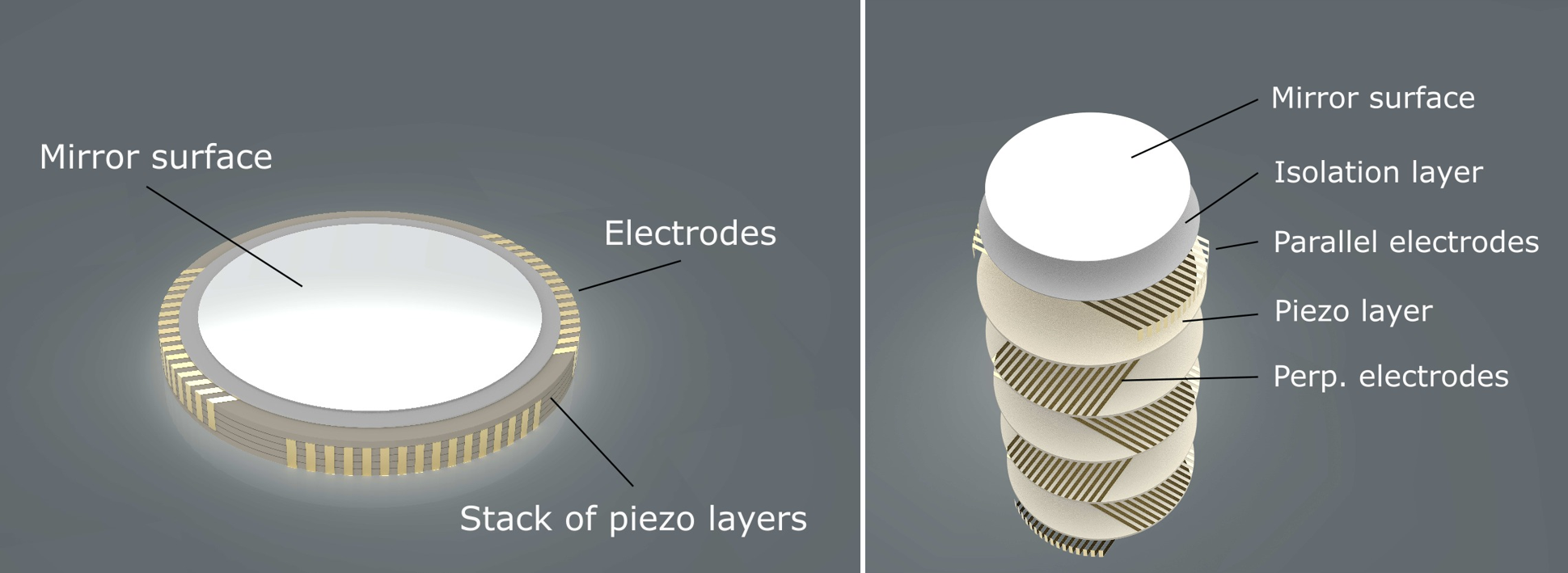}
    \caption{3D visualization of the mirror concept. The HDM consists of multilayered piezoelectric actuators which can deform the mirror surface by application of an electrical potential to the electrodes. Exploded view of the HDM with respective description of the individual components: mirror surface, isolation layer, parallel electrodes, piezo layers and perpendicular electrodes.}
    \label{fig:HDMconcept}
\end{figure}

The recently presented concept of a high pixel number deformable mirror utilizing piezoelectric hysteresis for stable shape configurations \cite{Huisman2018}, abbreviated hysteretic deformable mirror (HDM), demonstrates what we believe to the best of our knowledge is a new DM concept whose actuation mechanism consists of multilayered piezoelectric actuators with high hysteresis.  Figure \ref{fig:HDMconcept} shows a schematic illustration of the HDM. The high hysteresis of the newly developed piezoelectric material guarantees a remnant deformation of the mirror surface after the input has been removed. This property enables the combination of a simple electrode layout to define actuators at the intersections and use multiplexing to address those. The control of the remnant of a single actuator is presented in \cite{Marco2020-2}. 

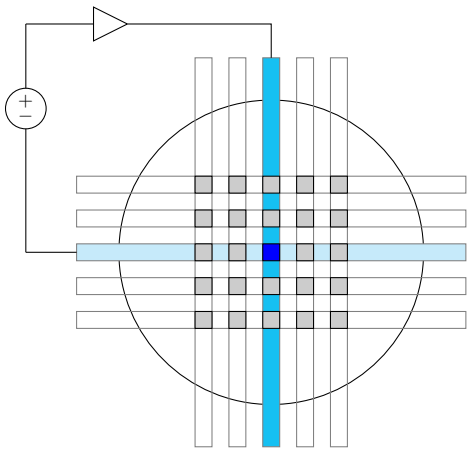
\begin{figure}[htbp]
    \centering
    \begin{tikzpicture}[scale=0.9]
    \draw[] (1.125,1.125) circle (2.25cm);
    \filldraw[fill=cyan!20!white!, draw=gray] (-1.75,1) rectangle (4,1.25);
    \draw[gray] (-1.75,0.5) rectangle (4,0.75);
    \draw[gray] (-1.75,0) rectangle (4,0.25);
    \draw[gray] (-1.75,1.5) rectangle (4,1.75);
    \draw[gray] (-1.75,2) rectangle (4,2.25);
    \draw[gray] (0,4) rectangle (0.25,-1.75);
    \draw[gray] (0.5,4) rectangle (0.75,-1.75);
    \filldraw[fill=cyan!70, draw=gray] (1,4) rectangle (1.25,-1.75);
    
    \draw[gray] (1.5,4) rectangle (1.75,-1.75);
    \draw[gray] (2,4) rectangle (2.25,-1.75); 
    
    \filldraw[fill=black!20!white] (0,0) rectangle (0.25,0.25); 
    \filldraw[fill=black!20!white] (0.5,0) rectangle (0.75,0.25);
    \filldraw[fill=black!20!white] (1,0) rectangle (1.25,0.25);
    \filldraw[fill=black!20!white] (1.5,0) rectangle (1.75,0.25);
    \filldraw[fill=black!20!white] (2,0) rectangle (2.25,0.25);
    
    \draw[fill=black!20!white] (0,0.5) rectangle (0.25,0.75);  
    \filldraw[fill=black!20!white] (0.5,0.5) rectangle (0.75,0.75);
    \filldraw[fill=black!20!white] (1,0.5) rectangle (1.25,0.75);
    \filldraw[fill=black!20!white] (1.5,0.5) rectangle (1.75,0.75);
    \draw[fill=black!20!white] (2,0.5) rectangle (2.25,0.75);
    
    \filldraw[fill=black!20!white] (0,1) rectangle (0.25,1.25);  
    \filldraw[fill=black!20!white] (0.5,1) rectangle (0.75,1.25);
    \filldraw[fill=blue, draw=black] (1,1) rectangle (1.25,1.25);

    \filldraw[fill=black!20!white] (1.5,1) rectangle (1.75,1.25);
    \filldraw[fill=black!20!white] (2,1) rectangle (2.25,1.25);
    
    \draw[fill=black!20!white] (0,1.5) rectangle (0.25,1.75); 
    \filldraw[fill=black!20!white] (0.5,1.5) rectangle (0.75,1.75);
    \filldraw[fill=black!20!white] (1,1.5) rectangle (1.25,1.75);
    \filldraw[fill=black!20!white] (1.5,1.5) rectangle (1.75,1.75);
    \draw[fill=black!20!white] (2,1.5) rectangle (2.25,1.75);
    
    \filldraw[fill=black!20!white] (0,2) rectangle (0.25,2.25); 
    \filldraw[fill=black!20!white] (0.5,2) rectangle (0.75,2.25);
    \filldraw[fill=black!20!white] (1,2) rectangle (1.25,2.25);
    \filldraw[fill=black!20!white] (1.5,2) rectangle (1.75,2.25);
    \filldraw[fill=black!20!white] (2,2) rectangle (2.25,2.25);
    
    %electrical circuit
    \draw[] (-1.75, 1.125) -- (-2.5,1.125);
    \draw[] (-2.5,1.125) -- (-2.5,2.95);
    \draw (-2.5,3.25) circle (0.3cm) node {$\substack{+ \\ -}$};
    \draw[] (-2.5,3.55) -- (-2.5, 4.5);
    \draw[] (-2.5,4.5) -- (-1.5,4.5);
    %triangle
    \draw[] (-1,4.5) -- (-1.5,4.75) -- (-1.5,4.25) -- (-1,4.5);
    \draw[] (-1,4.5) -- (1.125,4.5) -- (1.125,4);
        \end{tikzpicture}
    \caption{Conceptual electrode layout of the hysteretic deformable mirror from the top view. As an example in the illustration, the central actuator is actuated by application of an electrical potential to center top and bottom electrode which are visualized in blue while the other (not activated) actuators are represented in gray.}
    \label{fig:HDMelectrodes}
\end{figure}

The HDM focuses on applications where typically slowly varying disturbances ($>$ 1 Hz) must be corrected with extremely high accuracy ($<$ nm) and spatial frequency content, as it is the case for LUVOIR. Due to the HDM's design and working principle it is possible to employ a large number of actuators (128$\times$128) on an approximate surface area of \SI{900}{\square\milli\metre} and reach a high-resolution accuracy in correcting wavefront aberrations. In addition, it benefits from time-division multiplexing which reduces the number of wires needed to connect and address the actuators. Subsequently, the HDM provides a very simple electrode layout, as illustrated in Figure \ref{fig:HDMelectrodes}. The top and bottom electrodes are rotated by \ang{90} to form intersecting areas of the electrodes presenting the actuators. The actuation is bundled by sharing the same electrodes for actuators along a line. The voltage is transmitted over a shared top electrode while the corresponding bottom electrode for the desired actuator is grounded.

Motivated by this novel concept, we present the modeling and analysis of a mirror's facesheet that is subjected to the key characteristics of the HDM including a high actuator density and an interconnection layout. The mirror is described with a mechanical model to show the relation between the facesheet deflection and the pressures applied by the actuators. We follow the approach presented by Claflin and Bareket \cite{Claflin1986} in assuming that the deflection is governed by Poisson's equation. To guarantee a high accuracy in modeling, we incorporate the particular arrangement of the electrodes in the HDM into the solution to Poisson's equation and present the analytical solutions for the parametrization of squared electrodes as one of our contributions. In addition, an actuator model is introduced incorporating that the actuator’s pressure is a function of hysteresis modeled by the Preisach operator. Based on this, we can compute the required pressures to fit several Zernike polynomials \cite{Lakshminarayanan2011}, which are the preferred representation for light wavefront aberrations in adaptive optical systems. The simulation is performed for low actuator numbers to demonstrate the calculation method with the given conditions, and high actuator numbers which will allow a high spatial frequency wavefront correction. The results including the method's accuracy and limits of applicability are discussed.

The paper has four sections. Section \ref{sec:platemodel} presents the semi-analytical plate model to calculate the facesheet deflection caused by a high density array with square pressure planes of the actuators interacting with the facesheet. Section \ref{sec:results} describes the least-square fitting to determine optimal actuator pressures for representing wavefront aberrations and presents simulation results for a 5$\times$5 as well as a 129$\times$129 actuator array. Finally, results are discussed and the conclusions are given in Section \ref{sec:conclusions}.

\section{Semi-analytical plate model}\label{sec:platemodel}
An influence function defines the characteristic shape of the mirror surface corresponding to the deformation caused by one actuator. Several methods currently exist for modeling these influence functions of continuous facesheet mirrors. Besides the usage of Gauss functions and splines \cite{Huang2008,Tyson2000,Tyson2004}, or biharmonic plate equation \cite{Hamelinck2010}, influence functions can be modeled by application of Kirchhoff or van K\'{a}rm\'{a}n theory \cite{Grosso1977,Arnold1995,Ravensbergen2009} for plate deformations smaller than the plate thickness. Methods using the thin plate theory to calculate influence functions for real time computation for specific mirror geometries are given in \cite{Arnold1996,Arnold1997}. Furthermore, models based on the Kirchhoff plate model, for example, include assumptions for actuator forces that either presuppose the exerted force as point load or approximated electrode areas with constantly distributed loads as well as boundary conditions presenting circularly clamped DMs \cite{Bush2004,Claflin1986} or a free outer edge \cite{Menikoff1991}. Next to these modeling approaches which mainly consider the static characteristics, detailed review and analysis of DM's dynamic properties for control purposes can be found in \cite{Ruppel2012}. 

To determine the influence functions as precisely as possible with static characteristics, it is necessary to define the interaction areas according to their actual shapes. Given the concept of the HDM, the electrodes have an interconnection layout creating pressure planes lying under a thin circular facesheet. Since the upper electrodes are the closest layer to the facesheet besides the comparable thin isolation layer, which is neglected for simplicity of our modeling, we idealized these pressure planes as squares. The actuators are separated by a specified distance. To describe the surface displacement, it is necessary to integrate over the area of each pressure plane. Therefore, each plane is separated into several areas which can be described by a coordinate transformation using Cartesian coordinates as well as the radial and angular limits. It is assumed that the thickness of electrodes can be neglected and the piezoelectric actuators modeled as springs in parallel to a force source over an area which creates pressure on the facesheet.

\subsection{Determination of influence matrix}
We consider the Poisson equation \cite{Morse1953}

\begin{equation}
    \nabla^2 z=-\frac{q}{T}
\end{equation}
which governs the relation between small surface displacements $z$ of a thin facesheet with surface tension $T$ generated by an exerted pressure $q$. The solution to Poisson's equation in polar coordinates $(r,\phi)$ can be given by

\begin{equation}\label{eq:PoissonPolar}
z(r,\phi, \bar{r}, \bar{\phi})= C \iint_{A} \mathcal{F} (r,\phi,\bar{r},\bar{\phi}(\bar{r})) \,q(\bar{r},\bar{\phi}) \,d\bar{\phi}\,d \bar{r}
\end{equation}
with 

\begin{equation}
\begin{split}
A=\{(\bar{r},\bar{\phi}) | \phi_{1}(\bar{r}) \leq \bar{\phi} \leq \phi_{2}(\bar{r}), 0 \leq \bar{r} \leq 1\}
\end{split}
\end{equation}
where $z(r,\phi)$  is the out-of-plane displacement of the thin facesheet, $(\bar{r}, \bar{\phi})$ are the integration variables, $q(\bar{r},\bar{\phi})$ are the distributed forces over the particular electrode area, and constant $C=a^2/T$ contains the relation between the facesheet radius $a$ and the surface tension for normalization of the function $\mathcal{F}$ to unity. Edge deflection and slopes are both equal to zero. Furthermore,  $\mathcal{F}$ is defined as

\begin{equation}
\mathcal{F}(r,\phi,\bar{r},\bar{\phi}(\bar{r})) = \left\{\begin{array}{ll}
f_{1}(r,\phi,\bar{r},\bar{\phi}(\bar{r})) & \text{if} \quad 0<\bar{r}<r  \\
f_{2}(r,\phi,\bar{r},\bar{\phi}(\bar{r})) & \text{if} \quad r<\bar{r}<1. \\
\end{array}
\right. 
\end{equation}
The resulting deflection will be the integral of $z(r,\phi,\bar{r},\bar{\phi})$ over the area $A$ of the facesheet

\begin{equation}
z(r,\phi)= 
\frac{q(r,\phi) a^2}{2\pi T}  \int_{0}^{1} \int_{0}^{2\pi} \mathcal{F}(r,\phi,\bar{r},\bar{\phi}(\bar{r})) \,d\bar{\phi} \,d \bar{r} 
\end{equation}
assuming that $q(\bar{r},\bar{\phi})$ is a piecewise constant function on $R_1<r<R_2$ which gives $q(r,\phi)$. Note that $R_1$ and $R_2$ designate the smallest and greatest radius for describing the electrodes, respectively. 

Following the approach of Claflin and Bareket \cite{Claflin1986}, the equation for calculation of the surface deflection on a specific point on the clamped facesheet can be formulated as

\begin{equation} \label{eq:z}
    z(r,\phi)=\sum_{j=1}^{N_{e}}\mathcal{M}_{(r,\phi)j} q_{(r,\phi)j}
\end{equation}
where $\mathcal{M}$ represents the coefficients derived from the solutions of the Poisson equation, $q_{(r,\phi)j}$ are piecewise constant pressures exerted on the respective $j$-th electrode, and $N_{e}$ is the total number of electrodes.

The exact shape of an electrode is defined via a coordinate transformation. This allows us to implement the information later to the Poisson's equation (Eq. \ref{eq:PoissonPolar}) and find a solution. The electrode is split into parts based on areas of radial limits. These radial limits are used to implement the transformation from Cartesian to polar coordinates. Thus, $\phi$ depends on $r$. 
For convenience, the integration with respect to $\phi$ is performed first, and results in

\begin{equation}
\begin{split}
    z(r,\phi) = \frac{q(r,\phi)}{2\pi} C \Big\{ 
     -&\ln{(r)} \int_{0}^{r} \bar{r} (\phi_{2}(\bar{r}) - \phi_{1}(\bar{r}))  \\
     -& \bar{r} \sum_{n=1}^{\infty} \frac{1}{n^2} \Big(\Big(\bar{r} r\Big)^n - \Big(\frac{\bar{r}}{r}\Big)^n \Big) \\ 
    \times [\sin(n(\phi_{2}(\bar{r})-&\phi))-\sin(n(\phi_{1}(\bar{r})-\phi))] \,d \bar{r} \\
    +&\int_{r}^{1} \bar{r} \ln{\Big(\frac{1}{\bar{r}}\Big)} (\phi_{2}(\bar{r})-\phi_{1}(\bar{r})) \\
    -&\bar{r} \sum_{n=1}^{\infty} \frac{1}{n^2} \Big(\Big(\bar{r} r \Big)^n - \Big(\frac{r}{\bar{r}}\Big)^n\Big) \\ 
    \times [\sin(n(\phi_{2}(\bar{r})-&\phi))-\sin(n(\phi_{1}(\bar{r})-\phi))] \,d \bar{r} \Big\}.
\end{split} 
\end{equation}

\begin{figure}
    \centering
    \begin{tikzpicture}[scale=0.9]
    \filldraw[fill=black!60!white, draw=black] (0,0) rectangle (0.25,0.25); 
    \filldraw[fill=black!5!white, draw=black] (0.5,0) rectangle (0.75,0.25);
    \filldraw[fill=black!20!white, draw=black] (1,0) rectangle (1.25,0.25);
    \filldraw[fill=black!5!white, draw=black] (1.5,0) rectangle (1.75,0.25);
    \filldraw[fill=black!60!white, draw=black] (2,0) rectangle (2.25,0.25);
    
    \draw[pattern=north east lines] (0,0.5) rectangle (0.25,0.75);  
    \filldraw[fill=black!60!white, draw=black] (0.5,0.5) rectangle (0.75,0.75);
    \filldraw[fill=black!20!white, draw=black] (1,0.5) rectangle (1.25,0.75);
    \filldraw[fill=black!60!white, draw=black] (1.5,0.5) rectangle (1.75,0.75);
    \draw[pattern=north east lines] (2,0.5) rectangle (2.25,0.75);
    
    \filldraw[fill=black!20!white, draw=black] (0,1) rectangle (0.25,1.25);  
    \filldraw[fill=black!20!white, draw=black] (0.5,1) rectangle (0.75,1.25);
    \filldraw[fill=black, draw=black] (1,1) rectangle (1.25,1.25);
    \draw[] (1.125,1.125) circle (2.25cm);
    \filldraw[fill=black!20!white, draw=black] (1.5,1) rectangle (1.75,1.25);
    \filldraw[fill=black!20!white, draw=black] (2,1) rectangle (2.25,1.25);
    
    \draw[pattern=north east lines] (0,1.5) rectangle (0.25,1.75); 
    \filldraw[fill=black!60!white, draw=black] (0.5,1.5) rectangle (0.75,1.75);
    \filldraw[fill=black!20!white, draw=black] (1,1.5) rectangle (1.25,1.75);
    \filldraw[fill=black!60!white, draw=black] (1.5,1.5) rectangle (1.75,1.75);
    \draw[pattern=north east lines] (2,1.5) rectangle (2.25,1.75);
    
    \filldraw[fill=black!60!white, draw=black] (0,2) rectangle (0.25,2.25); 
    \filldraw[fill=black!5!white, draw=black] (0.5,2) rectangle (0.75,2.25);
    \filldraw[fill=black!20!white, draw=black] (1,2) rectangle (1.25,2.25);
    \filldraw[fill=black!5!white, draw=black] (1.5,2) rectangle (1.75,2.25);
    \filldraw[fill=black!60!white, draw=black] (2,2) rectangle (2.25,2.25);
    
    \filldraw[fill=black, draw=black] (4,2) rectangle (4.25,2.25);
    \node[] at (5,2.125) {Case 1}; %case1 
    \filldraw[fill=black!60!white, draw=black] (4,1.5) rectangle (4.25,1.75);
    \node[] at (5,1.625) {Case 2}; %case2
    \filldraw[fill=black!20!white, draw=black] (4,1) rectangle (4.25,1.25);
    \node[] at (5,1.125) {Case 3}; %case3 
    \filldraw[fill=black!5!white, draw=black] (4,0.5) rectangle (4.25,0.75);
    \node[] at (5,0.625) {Case 4}; %case4 
    \draw[pattern=north east lines] (4,0) rectangle (4.25,0.25); 
    \node[] at (5,0.125) {Case 5}; %case5 
    \draw[] (3.875,-0.125) rectangle (5.625,2.375);
        \end{tikzpicture}
    \caption{Conceptual top view of a 5$\times$5 actuator array to illustrate the classification of cases which arises from the geometric description and modeling method.}
    \label{fig:topview_cases}
\end{figure}
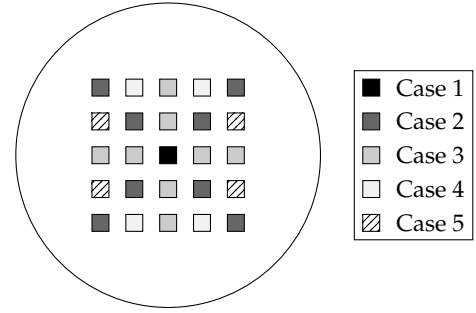

The introduced coordinate transformation is inserted and the integration with respect to $r$ is solved as a function of the position of the electrodes. We define five cases according to the actuator position (Figure \ref{fig:topview_cases}), as follows: \textit{Case 1,} central actuator; \textit{Case 2,} diagonal actuators; \textit{Case 3,} midline actuators; \textit{Case 4,} actuators above the diagonals; and \textit{Case 5,} actuators below the diagonal. The definition of each radial limit can be found in Table \ref{tab:radiallimits}, and is visualized in Figure \ref{fig:topviewcase1}-\ref{fig:topviewcase5} together with a respective pressure plane.

The detailed summary of the calculation of the coefficients resulting from the solution to Poisson's equation can be found in Appendices \ref{sec:resformulas} to \ref{a:Case5}.

\subsubsection{Case 1 - central actuator}
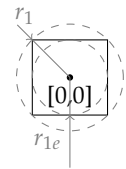
\begin{figure}[h!]
    \centering
 \begin{tikzpicture}
 \filldraw (0,0) node [below] {[0,0]} circle (1pt);
\draw (-0.5,-0.5) rectangle (0.5,0.5);
\draw[ gray] (0,0) -- (-0.5,0.5);
\draw[ gray,<-] (-0.5,0.5) -- (-0.7,0.7) node [midway, above] (TextNode) {$r_1$};
\draw[gray] (0,0) -- (0,-0.5);
\draw[gray,<-] (0,-0.5) -- (0,-1.2) node[midway, below,left] (TextNode) {$r_{1e}$};
\draw[dashed,gray] (0,0) circle (0.71);
\draw[dashed,gray] (0,0) circle (0.5);
\end{tikzpicture}  
    \caption{Definitions of the radial limits to describe \textit{Case 1,} central actuator.}
    \label{fig:topviewcase1}
\end{figure}

The central actuator (visualized in Figure \ref{fig:topviewcase1}) was described by use of two radial limits, $r_{1}$ and $r_{1e}$, where $r_{1}$ denotes the radius measured from the center to the corner points and $r_{1e}$ denotes an extra radius measured from the center to the inner side length. The calculation of coefficients for this case can be found in Appendix \ref{a:Case1}.

\subsubsection{Case 2 - diagonal actuators}
\begin{figure}[h!]
    \centering
    \begin{tikzpicture}
\draw[dashdotted] (-1,0) -- (1,0);
\draw[dashdotted] (0,1) -- (0,-1);
\filldraw (0,0) node [below] {[0,0]} circle (1pt);
\draw (1.5,2.5) rectangle (2.5,1.5); 
\draw [gray,->] (0,0) -- node[sloped, anchor=center, xshift=4ex, above] {$r_1$} (1.5,1.5);
\draw[gray,dashed]  (1.5,3.1) to[bend left] (3.1,1.5);
\draw [gray,->] (0,0) -- node[sloped, anchor=center, above] {$r_4$} (3.1,1.5);
\draw[gray,dashed]  (1.5,2.5) to[bend left] (2.5,1.5);
\draw [gray,->] (0,0) -- node[sloped, anchor=center, above] {$r_2=r_3$} (1.5,2.5);
  \end{tikzpicture}
    \caption{Definitions of the radial limits to describe \textit{Case 2,} diagonal actuator.}
    \label{fig:topviewcase2}
\end{figure}
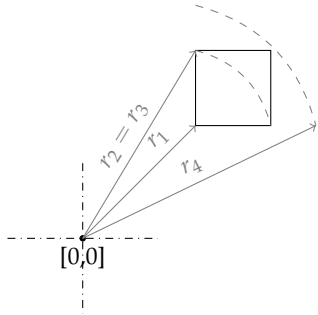

The actuators that lie on the diagonals (visualized in Figure \ref{fig:topviewcase2}) were described by use of three radial limits: $r_1$, $r_2=r_3$ and $r_4$. The numbering of the radii is systematically distributed according to the corner position. The calculation of coefficients for this case can be found in Appendix \ref{a:Case2}.

\subsubsection{Case 3 - midline actuators}
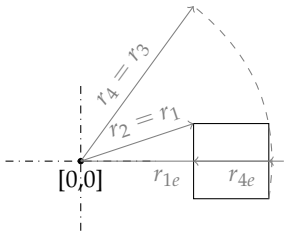
\begin{figure}[h!]
    \centering
    \begin{tikzpicture}
\draw[dashdotted] (-1,0) -- (1,0);
\draw[dashdotted] (0,1) -- (0,-1);
\filldraw (0,0) node [below] {[0,0]} circle (1pt);
\draw (1.5,-0.5) rectangle (2.5,0.5); 
\draw [gray,->] (0,0) -- node[sloped,anchor=center, above, xshift=.6em] {$r_2=r_1$} (1.5,0.5);
\draw[gray,dashed]  (2.5,-0.5) to[bend right] (1.5,2.0616);
\draw [gray,->] (0,0) -- node[sloped, anchor=center, above] {$r_4=r_3$} (1.5,2.0616);
\draw[gray] (0,0) -- (1.5,0) node[midway,below, xshift=2.7ex] (TextNode) {$r_{1e}$};
\draw[gray,<-] (1.5,0) --(2.5,0);
\draw[gray] (1.5,0) -- (2.5,0) node[midway,below, xshift=1ex] (TextNode) {$r_{4e}$};
\draw[gray,<-] (2.5,0) -- (2.7,0);
  \end{tikzpicture}
    \caption{Definitions of the radial limits to describe \textit{Case 3,} midline actuator.}
    \label{fig:topviewcase3}
\end{figure}

The actuators that lie on the midlines (visualized in Figure \ref{fig:topviewcase3}) were described by use of four radial limits: $r_{1e}$, $r_1$, $r_4$ and $r_{4e}$. $r_{1}$ and $r_{4}$ denote radii measured from the center to certain corner points and $r_{1e}$ and $r_{4e}$ denote extra radii indicating inner side lengths. The calculation of coefficients for this case can be found in Appendix \ref{a:Case3}.

\subsubsection{Case 4 - actuators above the diagonal}
\begin{figure}[h!]
    \centering
    \begin{tikzpicture}
\draw[dashdotted] (-1,0) -- (1,0);
\draw[dashdotted] (0,1) -- (0,-1);
\draw[dashdotted] (1,1) -- (-1,-1);
\draw[dashdotted] (-1,1) -- (1,-1);
\filldraw (0,0) node [below] {[0,0]} circle (1pt);
    \draw (1.5,3.5) rectangle (2.5,4.5);
    \draw[gray,->] (0,0) -- node[sloped,anchor=center,above,xshift=3.8em] (TextNode) {$r_1$} (1.5,3.5) ;
    \draw[gray,->] (0,0) -- node[sloped,anchor=center, above,xshift=3.1em] (TextNode) {$r_4$} (2.5,4.5) ;
    \draw[gray,dashed] (3.08,3.1)  arc[radius = 4.30116, start angle= 50, end angle= 75] node[] (TextNode) {};
    \draw[gray,->] (0,0) -- node[sloped,anchor=center,above] (TextNode) {$r_3$} (3.08, 3.1);
    \draw[gray,dashed] (2.5,4)  arc[radius = 3.4, start angle= 53, end angle= 76] node[] (TextNode) {};
    \draw[gray,->] (0,0) -- node[sloped,anchor=center, above,xshift=0.7em] (TextNode) {$r_2$} (1.2,4.6);
    \end{tikzpicture}
    \caption{Definitions of the radial limits to describe \textit{Case 4,} actuators above the diagonal.}
    \label{fig:topviewcase4}
\end{figure}
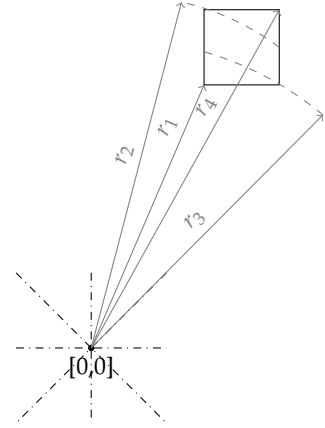

The actuators that lie above the diagonals (visualized in Figure \ref{fig:topviewcase4}) were described by use of four radial limits: $r_1$, $r_3$, $r_2$ and $r_4$. The numbering of the radii is systematically distributed according to the corner position. The calculation of coefficients for this case can be found in Appendix \ref{a:Case4}.

\subsubsection{Case 5 - actuators below the diagonal}
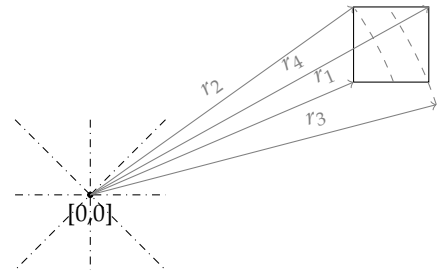
\begin{figure}[h!]
    \centering
\begin{tikzpicture}
\draw[dashdotted] (-1,0) -- (1,0);
\draw[dashdotted] (0,1) -- (0,-1);
\draw[dashdotted] (1,1) -- (-1,-1);
\draw[dashdotted] (-1,1) -- (1,-1);
\filldraw (0,0) node [below] {[0,0]} circle (1pt);
    \draw (3.5,1.5) rectangle (4.5,2.5); 
    \draw[gray,->] (0,0) -- node[sloped,anchor=center,above,xshift=5em] (TextNode) {$r_1$} (3.5,1.5);
    \draw[gray,->] (0,0) -- node[sloped,anchor=center,above, xshift=2em] (TextNode) {$r_4$} (4.5,2.5);
    \draw[gray,dashed] (4.6,1.2)  arc[radius = 5.3, start angle= 18.4349, end angle= 34] node[left,above, xshift=5ex, yshift=-7ex] (TextNode) {};
    \draw[gray,dashed] (4.031162,1.5)  arc[radius = 4.301162, start angle= 20.41 , end angle= 35] node[left,above, xshift=0ex, yshift=0ex] (TextNode) {};
    \draw[gray,->] (0,0) --  node[sloped,anchor=center,above] (TextNode) {$r_2$} (3.5,2.5);
    \draw[gray,->] (0,0) -- node[sloped,anchor=center,above,xshift=2.5em] (TextNode) {$r_3$} (4.6,1.2);
   \end{tikzpicture}
    \caption{Definitions of the radial limits to describe \textit{Case 5,} actuators below the diagonal.}
    \label{fig:topviewcase5}
\end{figure}

The actuators that lie below the diagonals (visualized in Figure \ref{fig:topviewcase5}) were described by use of four radial limits: $r_1$, $r_3$, $r_3$ and $r_4$. The numbering of the radii is systematically distributed according to the corner position. The calculation of coefficients for this case can be found in Appendix \ref{a:Case5}.

%case 1, case 2, case 3, case 4, case 5 : 
\begin{table}[h!]
\caption{\bf Definition of radial limits for splitting the electrode areas.}
\label{tab:radiallimits}
\centering
\resizebox{\columnwidth}{!}{
\begin{tabular}{@{}l|l|l|l|l@{}}
\textbf{Case 1}  & \textbf{Case 2} & \textbf{Case 3} & \textbf{Case 4} & \textbf{Case 5} \\ \midrule
$\bar{r}=0$ & $\bar{r}\leq r_1$ & $\bar{r}\leq r_{1e}$  & $\bar{r} \leq r_1$ &  $\bar{r} \leq r_{1}$          \\
$\bar{r} \geq r_{1}$  & $\bar{r} \geq r_4$       & $\bar{r} \geq r_4$      & $\bar{r} \geq r_4$        &  $\bar{r} \geq r_4$        \\
$0<\bar{r}\leq r_{1e} $  & $r_1<\bar{r} \leq r_2$   & $r_{1e}<\bar{r} \leq r_1$ & $r_1<\bar{r} \leq r_3$        &  $r_1<\bar{r} \leq r_2$         \\
$r_{1e} < \bar{r} < r_{1}$& $r_2 < \bar{r} <r_4$     & $r_1<\bar{r}\leq r_{4e}$ & $r_3 < \bar{r} \leq r_2$        & $r_2<\bar{r} \leq r_3$          \\
& & $r_{4e}<\bar{r}<r_4$   &   $r_2<\bar{r} <r_4$        &   $r_3<\bar{r}<r_4$     \\ 
\end{tabular}%
}
\end{table}

\subsection{Actuator model}
\begin{figure}[h!]
    \centering
\begin{tikzpicture}[scale=0.5]
\draw[line width=0.5mm, scale=2.25, xshift=-0.3em] (-0.2,0.6) sin (1,1 ) cos(2.2,0.6);
\node[] (a) at (2,-2) {};
\draw[decoration={aspect=0.3, segment length=1.5mm, amplitude=1.25mm,coil},decorate] (2,1.5) -- (a); 
\fill [pattern = north east lines] (-1,-1.9) rectangle (5.5,-2.1);
\draw[thick] (-1,-1.9) -- (5.5,-1.9);
\draw (2,1.5) -- (2,2.15);
\draw[line width=0.25mm] (1.75,2.15) -- (2.25,2.15);
\draw (2,1.75) -- (2.75,1.75);
\draw (2.75,1.75) -- (2.75,0.2);
\draw (2.75,-0.2) circle (0.4cm) node {$\Phi$};
\draw (2.75,-0.6) -- (2.75,-1.9);
\node[] at (1.4,-0.2) {$k$};
\node[] at (7,1.1) {mirror facesheet};
\node[] at (6,-1.4) {actuator model};
\draw (2,-1.7) -- (2,-1.9);
\end{tikzpicture}
    \caption{Simplified actuator model, modeled by a stiffness $k$ in parallel to a force source over an area $\Phi$ acting on the mirror facesheet. }
    \label{fig:actuator}
\end{figure}
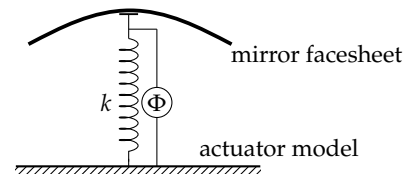

The actuators become coupled through the stiffness of the facesheet. Usually, DMs profit by low inter-actuator coupling, denoting the mechanical coupling between neighboring actuators, which improves the surface accuracy. If significant inter-actuator coupling is present, this needs to be considered in the modeling and control processes \cite{Hardy1998}. 
Here, we introduce the model of actuators based on two components, which correspond to a spring in parallel with a force source (Figure \ref{fig:actuator}). 

\begin{figure}[htbp]
    \centering
    \includegraphics[scale=0.4]{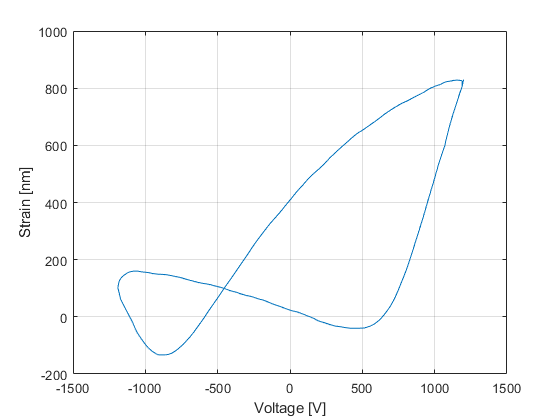}
    \caption{Asymmetric butterfly hysteresis loop with remnant deformation, the measured data of which was collected from previous material tests. The axial displacement was measured while a certain voltage was applied.}
    \label{fig:butterflyloop}
\end{figure}

The pressure term $q(r,\phi)$ can  be split so that it captures both components in terms of stiffness and force source over an area. Consequently, the relation from Equation \eqref{eq:z} may be described by

\begin{equation}    z(r,\phi)=\sum_{j=1}^{N_{e}}\mathcal{M}_{(r,\phi)j} \bigg(\Phi_{Pj}(V)-k_{j}\widetilde{z_{j}}\bigg)
\end{equation}
with 

\begin{equation}
\Phi_{Pj}(V)= Y_{j} \Phi_{Tj}(V) 
\end{equation}
and 

\begin{equation}
\widetilde{z_{j}}:=\frac{\sum_{i \in Ej} z_{i}/n_e}{A_e}
\end{equation}
where $\Phi_{Pj}(V)$ denotes the Preisach operator capturing the highly nonlinear hysteresis of the actuators in regard to the total deformation in relation of the initial thickness dimension, the diagonal matrix containing the Young's modulus $Y_{j}$, the longitudinal elongations of the actuators $\Phi_{Tj}(V)$, the diagonal stiffness matrix containing the actuators' stiffness $k_{j}$, and the mean surface deflection above the respective electrode with area $A_e$ $\widetilde{z}_{j}$ calculated by means of $n_e$ surface displacement points $z_i$ on a specific position within the electrode area. It is assumed that all the actuators are identical and can exert an asymmetric butterfly loop as exemplary presented in Figure \ref{fig:butterflyloop}. A framework to model the electric-field dependence on the strain in piezoelectric materials purposely designed to exhibit loops with remnant deformation was presented by Jayawardhana et al. \cite{Bayu2018} based on the use of the Preisach operator. The complete formal definition of the Preisach operator is given in \cite{Marco2020}.

\section{Results and discussion}\label{sec:results}
\subsection{2D pattern for Influence Functions of \textit{Case} 1 - 5}
\begin{figure}
    \centering
    \includegraphics[trim=120 0 120 0,width=\columnwidth]{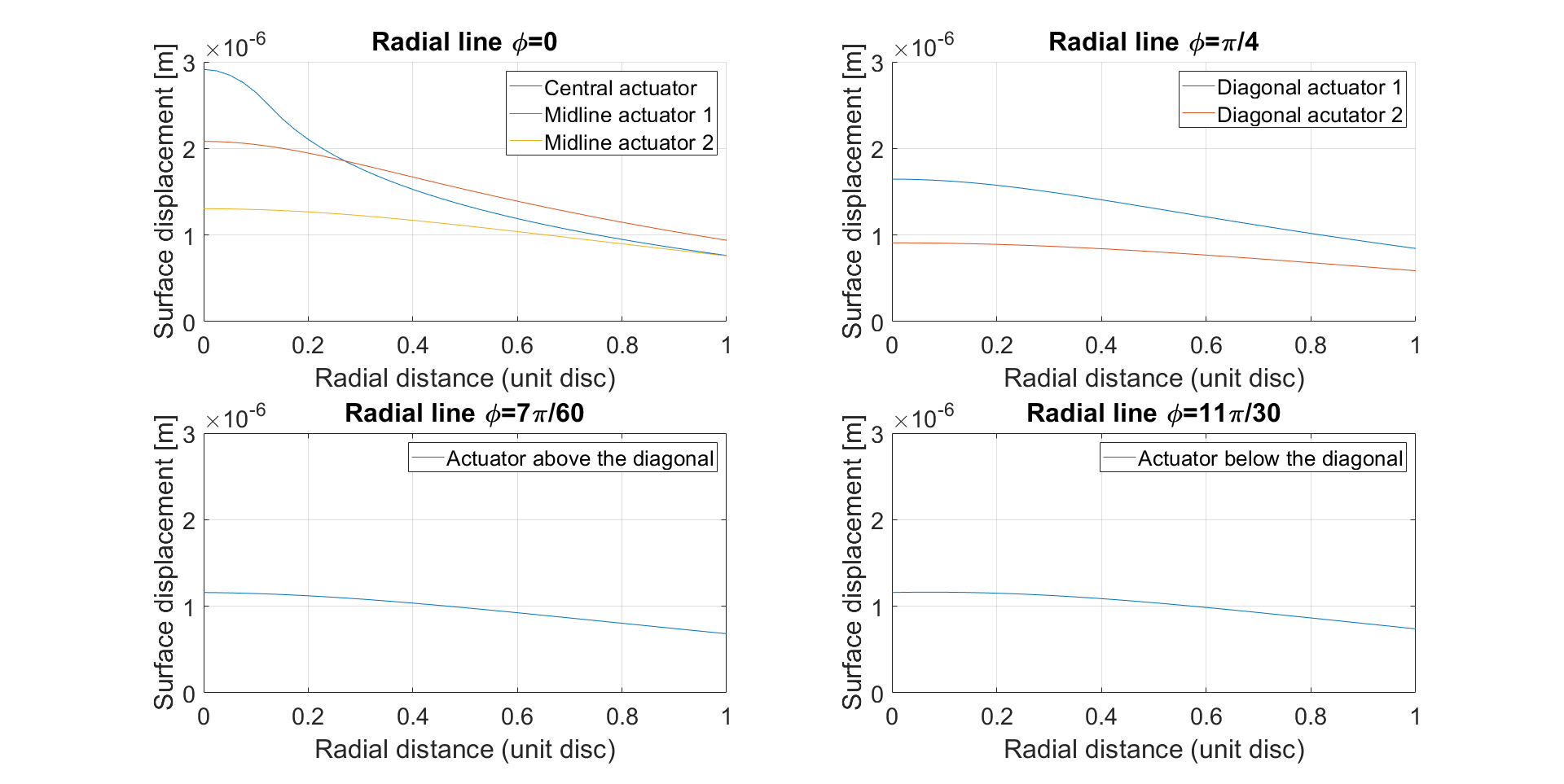}
    \caption{Influence functions in the optically active area for \textit{Case} 1 and \textit{Case} 3 (top, left),  \textit{Case} 2 (top, right),  \textit{Case} 4 (bottom, left) and  \textit{Case} 5 (bottom, right) plotted along the radial line $\phi$. Each actuator was addressed individually with a pressure of 0.01 \si{\newton\per\square\metre}.}
    \label{fig:2dpattern}
\end{figure}

The influence function of every case can be seen in Figure \ref{fig:2dpattern}. The actuators were individually addressed, and their arrangement corresponds to the 5$\times$5 actuator array that is exemplarily visualized in Figure \ref{fig:topview_cases}.

%\begin{figure}
%    \centering
%    \includegraphics[scale=0.4]{2Dpattern/patterncentermidline.png}
%    \caption{Influence function in the active area for \textit{Case} 1 - central actuator (blue), \textit{Case} 3 - midline actuators (orange and yellow) plotted along the radial line $\phi=0$. Each actuator was addressed individually with a pressure of 0.01 \si{\newton\per\square\metre}.}
%    \label{fig:2dpattern_12}
%\end{figure}

%\begin{figure}
%    \centering
%    \includegraphics[scale=0.4]{2Dpattern/patterndiagonalactuators.png}
%    \caption{Influence function in the active area for \textit{Case} 2 - diagonal actuators (blue and orange) plotted along the radial line $\phi=\pi/4$. Each actuator was addressed individually with a pressure of 0.01 \si{\newton\per\square\metre}.}
%    \label{fig:2dpattern_3}
%\end{figure}

%\begin{figure}
%    \centering
%    \includegraphics[scale=0.4]{2Dpattern/patternactuatorabovedia.png}
%    \caption{Influence function in the active area for \textit{Case} 4 - actuators above the diagonal plotted along the radial line $\phi=7\pi/60$. The actuator was addressed individually with a pressure of 0.01 \si{\newton\per\square\metre}.}
%    \label{fig:2dpattern_4}
%\end{figure}

%\begin{figure}
%    \centering
%    \includegraphics[scale=0.4]{2Dpattern/patternactuatorbelowdia.png}
%    \caption{Influence function in the active area for \textit{Case} 5 - actuators below the diagonal plotted along the radial line $\phi=11\pi/30$. The actuator was addressed individually with a pressure of 0.01 \si{\newton\per\square\metre}.}
%    \label{fig:2dpattern_5}
%\end{figure}

\subsection{Least-square fitting}
The preferred representation for light wavefront aberrations in adaptive optical systems is via Zernike polynomials. They are defined on a unit circle using polar coordinates $(r,\theta)$ as functions of azimuthal frequency $m$ and radial degrees $n$, where $m \leq n$ . The set of polynomials \cite{Lakshminarayanan2011} can be given by

\begin{equation}
\begin{split}
Z\substack{m\\n}(r,\theta)=R\substack{m\\n}(r)\cos (m\theta) & \quad \text{for} \quad m \geq 0 \\    
Z\substack{-m\\n}(r,\theta)=R\substack{m\\n}(r)\sin (m\theta) & \quad \text{for} \quad m<0
\end{split}
\end{equation}
where
\begin{equation}
R\substack{m\\n}(r)= \sum_{S=0}^{(n-m)/2} \frac{(-1)^S (n-S)!r^{n-2S})}{S![(n+m)/2-S]![(n-m)/2-S!]}.
\end{equation}

To calculate the required pressure terms to fit several Zernike polynomials, each displacement of a respective point on the facesheet which is defined by $(r,\phi)$ is fit to the corresponding point on Zernike polynomials. An over-determined set of equations is solved in the least-square sense resulting in 

%solution in terms of least-square fitting problem:
\begin{equation}
    \Phi=(\mathcal{M}^\intercal \mathcal{M})^{-1}\mathcal{M} (z_{d}+\mathcal{M} k\widetilde{z}_{d}).
\end{equation}
which aims at minimizing the root-mean-square deviation (RMSD) between the two quantities.

\subsection{Simulation results}
Using \textsc{Matlab} R2019a, a low density array with 5$\times$5 actuators and a high density array with 129$\times$129 actuators were generated. To decrease the computational effort in the latter case, the coefficient calculations were executed in parallel per 5 actuators and run in a compute cluster (Peregrine HPC cluster). For all experiments, we used a partition of two Intel Xeon E5 2680 v3 or v4 (2.50GHz or 2.40GHz respectively) CPUs with 5GB of memory. Thereby, the computational time was decreased to about 2h when all jobs ran in parallel.

To assess the mechanical model, a second simulation in Matlab was generated fitting the mirror surface to selected Zernike polynomials.
%as in Figure \ref{fig:ZP31} and \ref{fig:ZP04}. 
The procedure of this approach included three steps. The first step consisted of reducing the mirror surface to an optically active area due to the boundary conditions to circumvent an increasing fitting error caused by zero deflection at the clamped edge. Secondly, a mask was generated to match selected points of the Zernike polynomial disc plot to the surface points of the mirror. This mask was created with a partition in radial and angular coordinates according to $r_{0} < r_{1}< \dots < r_{(n-1)} < r_{n}$ with $r_{0}=0$ and $r_{n}=1$, and $\phi_{0} < \phi_{1} < \dots < \phi_{(n-1)} < \phi_{n}$ with $\phi_{0}=0$ and $\phi_{n}=2\pi$ respectively. In the third step, the RMSD of the estimator $z_{d}$ with respect to the actual surface deflection $z$ was calculated (\eqref{eq:RMSD}) to evaluate the mirror accuracy by

\begin{equation}\label{eq:RMSD}
    RMSD(z)=\sqrt{\frac{\sum_{x=1}^{X}(z_{x}-z_{d_{x}})^2}{x}}.
\end{equation}

\begin{figure}
    \centering
    \includegraphics[scale=0.5]{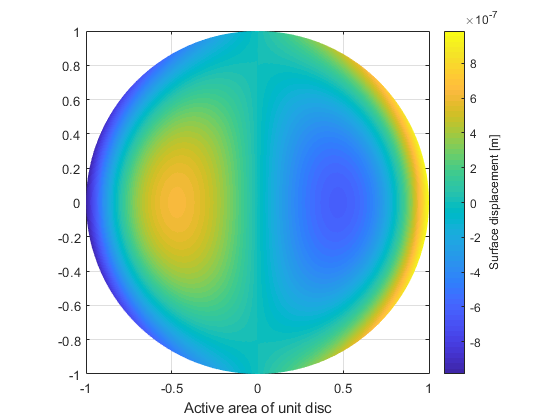}
    \caption{129$\times$129 actuator array fitted to Zernike polynomial $Z\protect\substack{1\\3}$ with a peak-to-valley amplitude of $1.972\protect\mu m$} in a graphic representation showing the active area of the mirror as unit disc with vertical colorbar giving the surface displacement.
    \label{fig:simulationresult1129x129}
\end{figure}

\begin{figure}
    \centering
    \includegraphics[scale=0.5]{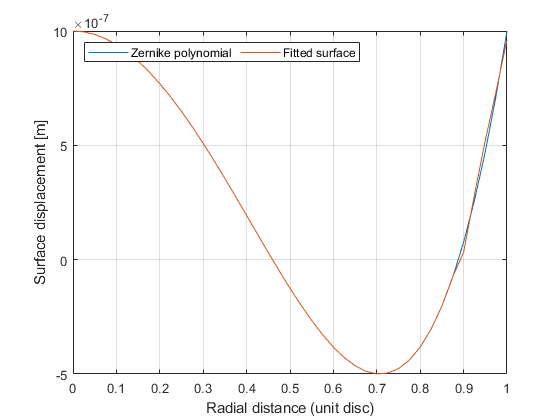}
    \caption{Surface displacement along the radial line $\phi=0$ for a 129$\times$129 actuator array fitted to Zernike polynomial $Z\protect\substack{0\\4}$ with a peak-to-valley amplitude of $1.199\protect\mu m$}.
    \label{fig:simulationresult4129x129}
\end{figure}

\begin{figure}
    \centering
    \includegraphics[scale=0.5]{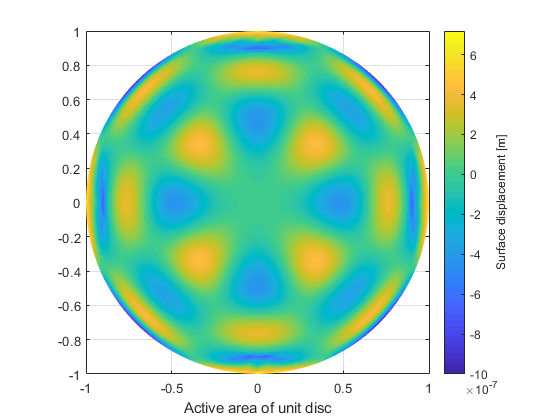}
    \caption{129$\times$129 actuator array fitted to Zernike polynomial $Z\protect\substack{4\\10}$ with a peak-to-valley amplitude of $1.718\protect\mu m$ in a graphic representation showing the active area of the mirror as unit disc with vertical colorbar giving the surface displacement.}
    \label{fig:simulationresult5129x129}
\end{figure}

\begin{figure}
    \centering
    \includegraphics[scale=0.5]{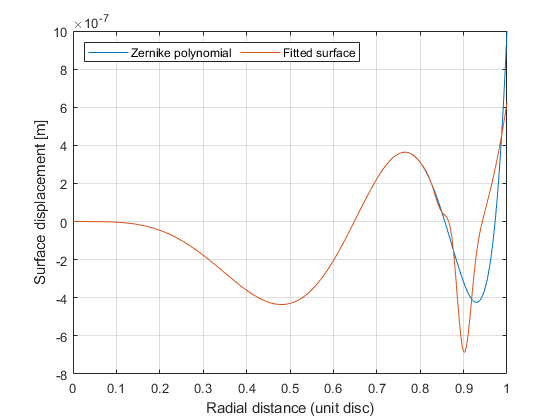}
    \caption{Surface displacement along the radial line $\phi=0$ for a 129$\times$129 actuator array fitted to Zernike polynomial $Z\protect\substack{4\\10}$ with a peak-to-valley amplitude of $1.718\protect\mu m$.}
    \label{fig:simulationresult6129x129}
\end{figure}

There were 4961 surface points selected based on the described partition over a diameter of 0.8 which corresponds to the active area. The facesheet surface tension amounted to 15 $\si{\newton\per\metre}$ and total mirror radius normalized to 1.
%Figures \ref{fig:simulationresult1} and \ref{fig:simulationresult2}, and Figures \ref{fig:simulationresult3} and \ref{fig:simulationresult4} show the results for a 5$\times$5 actuator array of fitting the mirror surface to selected Zernike polynomial examples. 
Figures \ref{fig:simulationresult1129x129}-% and \ref{fig:simulationresult2129x129}, and Figures \ref{fig:simulationresult3129x129} and
\ref{fig:simulationresult4129x129} show the results for a 129$\times$129 actuator array of fitting the mirror surface to lower order Zernike polynomials while Figure \ref{fig:simulationresult5129x129} and \ref{fig:simulationresult6129x129} show the result for a selected higher order polynomial.
%selected examples. 
Table \ref{tab:results} summarizes the RMSDs for the first 28 Zernike polynomials and selected higher order ones fitted with a peak-to-valley amplitude in the region of approximately 1.5 to 2$\protect\mu m$ with a low and high density array.

\begin{figure}
    \centering
    \includegraphics[scale=0.5]{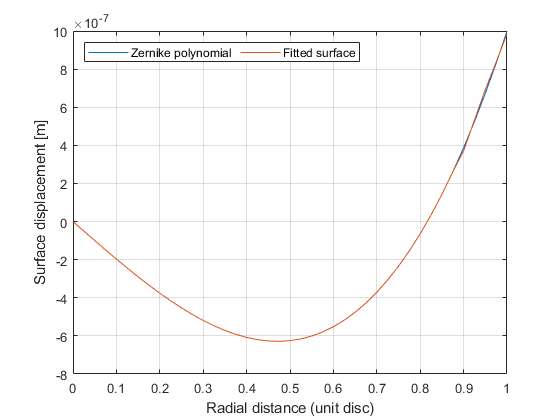}
    \caption{Surface displacement along the radial line $\phi=0$ for a 129$\times$129 actuator array fitted to Zernike polynomial $Z\protect\substack{1\\3}$ with a peak-to-valley amplitude of $1.972\protect\mu m$}.
    \label{fig:simulationresult2129x129}
\end{figure}

\begin{figure}
    \centering
    \includegraphics[scale=0.5]{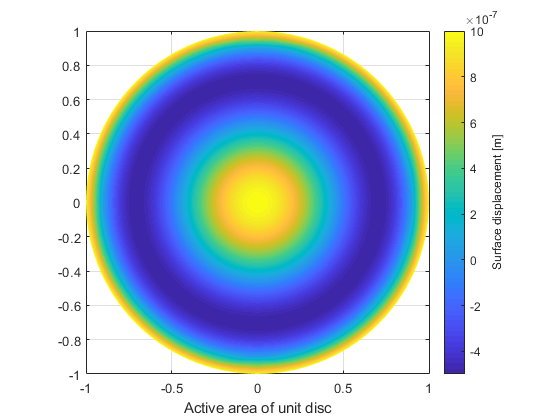}
    \caption{129$\times$129 actuator array fitted to Zernike polynomial $Z\protect\substack{0\\4}$ with a peak-to-valley amplitude of $1.199\protect\mu m$} in a graphic representation showing the active area of the mirror as unit disc with vertical colorbar giving the surface displacement.
    \label{fig:simulationresult3129x129}
\end{figure}

%%additional plots

\begin{table}  
\caption{\bf Summary of Root-Mean-Square Deviations (RMSDs) for selected Zernike polynomials (ZPs) with a 5$\times$5 (Low Density (LD)) and 129$\times$129 actuator array (High Density (HD)).}
    \label{tab:results}
    \centering
    %\begin{ruledtabular}
\begin{tabular}{@{}l|ll@{}}
\textbf{ZPs} &  \textbf{LD: RMSDs in [\%]}&  \textbf{HD: RMSDs in [\%]} \\ \midrule
$Z\substack{-1\\1}$ & 4.426 & 0.001433\\ 
$Z\substack{1\\1}$ &4.352& 0.001373  \\
$Z\substack{-2\\2}$ &3.806& 0.000298\\
$Z\substack{0\\2}$ &12.840& 0.065971\\
$Z\substack{2\\2}$ &9.498& 0.004775\\
$Z\substack{-3\\3}$ &10.138& 0.007609\\
$Z\substack{-1\\3}$ &13.937& 0.131942\\
$Z\substack{1\\3}$ &13.756& 0.134622 \\
$Z\substack{3\\3}$ &9.974& 0.007343\\
$Z\substack{-4\\4}$ &11.376& 0.005252\\
$Z\substack{-2\\4}$ &12.662& 0.082268\\
$Z\substack{0\\4}$ &25.917& 0.524411\\
$Z\substack{2\\4}$ &20.755& 0.409452\\
$Z\substack{4\\4}$ &13.069& 0.011954\\
$Z\substack{-5\\5}$ & 13.847 & 0.017128 \\
$Z\substack{-3\\5}$ & 18.909& 0.406282 \\
$Z\substack{-1\\5}$ & 18.838& 0.675500\\
$Z\substack{1\\5}$ & 18.674& 0.692096\\
$Z\substack{3\\5}$ & 18.717& 0.400501\\
$Z\substack{5\\5}$ & 13.649 & 0.016273\\
$Z\substack{-6\\6}$ &14.903 & 0.009467 \\
$Z\substack{-4\\6}$ & 20.851& 0.364835 \\
$Z\substack{-2\\6}$ &18.077 & 0.346565 \\
$Z\substack{0\\6}$ & 25.742 & 1.735849 \\
$Z\substack{2\\6}$ &17.145 & 1.670824\\
$Z\substack{4\\6}$ & 16.980& 0.729995 \\
$Z\substack{6\\6}$ & 14.790& 0.049802\\
$Z\substack{-10\\10}$ & 16.216& 0.087481\\
$Z\substack{-8\\10}$ & 18.114& 1.420633\\
$Z\substack{-6\\10}$ & 18.901 & 2.341578 \\
$Z\substack{-4\\10}$ & 20.592& 2.417903\\
$Z\substack{-2\\10}$ & 22.680& 1.451910\\
$Z\substack{0\\10}$ & 27.243 & 6.092243 \\
$Z\substack{2\\10}$ & 21.649& 6.225307 \\
$Z\substack{4\\10}$ & 18.076& 4.505920\\
$Z\substack{6\\10}$ & 18.868& 3.345968\\
$Z\substack{8\\10}$ & 17.221& 1.468534\\
$Z\substack{10\\10}$ & 17.303 & 0.271556\\
%$Z\substack{\\}$ & & \\
    \end{tabular}
   % \end{ruledtabular}
\end{table}

Considering the fitting results for a low density array with 5$\times$5 actuators (25 actuators in the active area), the intersection layout became clear and the positions of the few actuators play a major role for the final results. The fitting errors are between 3.8\% and 27.2\%. Comparing these results to a high density array with 129$\times$129 actuators (16073 actuators in the active area, 568 actuators outside), we observe that the RMSDs decrease drastically. With 129$\times$129 actuators, we have deviations between 0.000298\% and 6.23\% for the selected polynomials. For higher order polynomials corner effects are visible. Although the fitting errors for low and high density arrays behave in a similar manner with increasing degree of the polynomial, it is noticeable that with Zernike polynomials in the cosine phase the RMSD is slightly higher due to the square grid the HDM is characterized with. For further fitting improvement, the position of the square region, in which the actuators are created due to the interconnection of electrodes, might be adjusted with the active area to cover completely upper, lower, left and right corner of the unit disc.

\section{Conclusions}\label{sec:conclusions}
This study investigated the fundamental characteristics of actuator positions of high density arrays and presented a generalization of cases to calculate every actuator position of deformable mirrors for application in what we believe to the best of our knowledge is a novel hysteretic deformable mirror. Based on the introduced coordinate transformation while solving the Poisson equation, it was possible to model exactly the shape of the pressure planes and guarantee a more realistic description of the actuator influence functions. By calculating the coefficient matrix in a cluster, the computational time was decreased which presents a usable method for computations on deformable mirrors with high actuator densities. Furthermore, the mirror model includes the mechanical coupling between the actuators and the facesheet.
The presented results contribute to achieve a higher accuracy in modeling the actuator influence functions according to the actual properties of the DM, and therefore decrease fitting errors. It provides a framework on how to consider high actuator densities and calculate them in a reasonable way regarding actuator position case classification and computation time. 

%APPENDIX
\appendix
\section{Respective formulas}\label{sec:resformulas}
For actuators of the right side of the plate, the left corner of a pressure plane is denoted with $x_1$, the right corner with $x_2$, the lower corner with $y_1$ and the upper corner with $y_2$. The designation is mirrored with actuators on the left side of the plate. In general, it can be said that $\lvert x_1 \lvert \leq \lvert x_2 \lvert$ and $\lvert y_1 \lvert \leq \lvert y_2 \lvert$.

Table \ref{tab:coordinatetransformation} summarizes the definition of all radial limits with coordinate transformations for splitting the electrode areas. Symbols which are assigned to reoccurring formulas are listed in Table \ref{tab:symbols}.

\begin{table}[htbp]
\caption{\bf Definition of Radial Limits in Interval $I$ with Coordinate Transformation for Splitting the Electrode Areas.}
\label{tab:coordinatetransformation}
\centering
%\begin{ruledtabular}
\begin{tabular}{@{}lll@{}}
%\textbf{Case 1} && \\
\textbf{Case 1} & Boundaries      & Coordinate transformation    \\ \midrule
$I_1$ & $0 \leq \bar{r}\leq r_{1e}$ & $0< \bar{\phi}< 2\pi$ \\
$I_{21}$ & $r_{1e}<\bar{r}< r_{1}$& $\arccos{(x_{2}/\bar{r})}<\bar{\phi}<\arcsin{(y_{2}/\bar{r})}$ \\
$I_{22}$ &              & $\arcsin{(y_{2}/\bar{r})}<\bar{\phi}<\arccos{(x_{1}/\bar{r})}$ \\
$I_{23}$ &               & $\arccos{(x_{1}/\bar{r})}<\bar{\phi}<\arcsin{(y_{1}/\bar{r})}$ \\
$I_{24}$&& $\arcsin{(y_{1}/\bar{r})}<\bar{\phi}<\arccos{(x_{2}/\bar{r})}$ \\  \midrule
%\textbf{Case 2} && \\
\textbf{Case 2} & Boundaries  & Coordinate transformation \\  \midrule
$I_1$&$r_{1}<\bar{r}\leq r_{2}$& $\arcsin{(y_{1}/\bar{r})}<\bar{\phi}\leq\arccos{(x_{1}/\bar{r})}$ \\
$I_2$&$r_{2}<\bar{r}<r_{4}$& $\arccos{(x_{2}/\bar{r})}<\bar{\phi}<\arcsin{(y_{2}/\bar{r})}$ \\ \midrule
%\textbf{Case 3} && \\
\textbf{Case 3} & Boundaries  & Coordinate transformation      \\ \midrule
$I_{11}$&$r_{1e}<\bar{r}\leq r_{1}$& $0 < \bar{\phi} \leq \arccos{(x_{1}/\bar{r})}$               \\
$I_{12}$&                & $2\pi-\arccos{(x_{1}/\bar{r})}<\bar{\phi}<2\pi$                  \\
$I_{2}$&$r_{1}< \bar{r}\leq r_{4e}$& $\arcsin{(y_{1}/\bar{r})}<\bar{\phi}\leq\arcsin{(y_{2}/\bar{r})}$ \\
$I_{31}$&$r_{4e}<\bar{r}<r_{4}$& $\arccos{(x_{2}/\bar{r})}<\bar{\phi}<\arcsin{(y_{2}/\bar{r})}$ \\
$I_{32}$&                & $\arcsin{(y_{1}/\bar{r})}<\bar{\phi}<\arccos{(x_{2}/\bar{r})}$ \\ \midrule
%\textbf{Case 4} && \\
\textbf{Case 4} & Boundaries  & Coordinate transformation      \\ \midrule
$I_{1}$&$r_1 < \bar{r}\leq r_3$ & $\arcsin{(y_{1}/\bar{r})} <\bar{\phi} \leq \arccos{(x_{1}/\bar{r})}$ \\
$I_{2}$&$r_3 < \bar{r}\leq r_2$ & $\arccos{(x_{2}/\bar{r})} <\bar{\phi} \leq \arccos{(x_{1}/\bar{r})}$ \\
$I_{3}$&$r_2 < \bar{r} <r_4 $   & $\arccos{(x_{2}/\bar{r})} <\bar{\phi} < \arcsin{(y_{2}/\bar{r})}$ \\ \midrule
%\textbf{Case 5} && \\
\textbf{Case 5} & Boundaries  & Coordinate transformation      \\ \midrule
$I_{1}$&$r_1 < \bar{r} \leq r_2$ & $\arcsin{(y_{1}/\bar{r})} <\bar{\phi} \leq \arccos{(x_{1}/\bar{r})}$ \\
$I_{2}$&$r_2 < \bar{r} \leq r_3$ & $\arcsin{(y_{1}/\bar{r})} <\bar{\phi} \leq \arcsin{(y_{2}/\bar{r})}$ \\
$I_{3}$&$r_3 < \bar{r} <r_4$   & $\arccos{(x_{2}/\bar{r})} <\bar{\phi} < \arcsin{(y_{2}/\bar{r})}$ \\
\end{tabular}
%\end{ruledtabular}
\end{table}

\begin{table}[htbp]
\caption{\bf Assignment of Symbols to Reoccurring Formulas.}
\label{tab:symbols}
\centering
%\begin{ruledtabular}
\begin{tabular}{@{}ll|ll@{}}
\textbf{Symbol} & \textbf{Formula} & \textbf{Symbol} & \textbf{Formula}    \\ \midrule
$\kappa_{x_1}$ & $x_1\sqrt{1 - x_1^2/r^2}$ & $\kappa_{y_1}$  &$y_1\sqrt{1 - y_1^2/r^2}$ \\
$\kappa_{x_2}$ & $x_2\sqrt{1 - x_2^2/r^2}$ & $\kappa_{y_2}$ & $y_2\sqrt{1 - y_2^2/r^2}$ \\
$\epsilon_{x_1}$ & $x_1\sqrt{(r^2 - x_1^2)/r^2}$  & $\epsilon_{y_1}$ & $y_1\sqrt{(r^2 - y_1^2)/r^2}$ \\
$\epsilon_{x_2}$ & $x_2\sqrt{(r^2 - x_2^2)/r^2}$ & $\epsilon_{y_2}$ & $y_2\sqrt{(r^2 - y_2^2)/r^2}$\\
$\alpha_1$ & $\arccos(x_1/r)$ &$\alpha_2$&  $\arccos(x_2/r)$\\
$\beta_1$ & $\arcsin(y_1/r)$ &$\beta_2$& $\arcsin(y_2/r)$\\
$\gamma_1$ & $\arcsin(x_1/r)$ &$\gamma_2$ & $\arcsin(x_2/r)$ \\
\end{tabular}
%\end{ruledtabular}
\end{table}

\section{Numerical integration} \label{a:numint}
Here are the two sub-integrals that are solved numerically.  174 is the maximum number of $n$ terms required for convergence \cite{Claflin1986}, so
\begin{equation*}
\begin{split}
    f_{1n}:= \sum_{n=1}^{\infty} \int_{R_1}^{R_2} \frac{\bar{r}}{n^2} \left(\left(\bar{r}r\right)^n - \left(\frac{\bar{r}}{r}\right)^n \right) \\
    \times [\sin(n(\phi_{2}(\bar{r})-\phi))-\sin(n(\phi_{1}(\bar{r})-\phi))] \,d \bar{r} 
    \end{split}
\end{equation*}
\begin{equation*}
\begin{split}
    f_{2n}:=\sum_{n=1}^{\infty} \int_{R_1}^{R_2} \frac{\bar{r}}{n^2} \left(\left(\bar{r} r\right)^n - \left(\frac{r}{\bar{r}}\right)^n \right)  \\
    \times [\sin(n(\phi_{2}(\bar{r})-\phi))-\sin(n(\phi_{1}(\bar{r})-\phi))] \,d \bar{r}.
\end{split}
\end{equation*}

\section{Coefficient calculation in \textit{Case 1}}\label{a:Case1}
Here are the formulas for calculating the coefficients $\mathcal{M}$ for actuators that can be categorized in \textit{Case 1}:

\subsubsection{\texorpdfstring{$r_i=0$}{ri=0}}
\begin{equation*}
\begin{split}
\mathcal{M}=&(1/(2\pi)) \times ((f_{2(I_{1})}(r_{1e}))+(f_{2(I_{21})}(r_1)-f_{2(I_{21})}(r_{1e})) \\
&+(f_{2(I_{22})}(r_1)-f_{2(I_{22})}(r_{1e})) 
+(f_{2(I_{23})}(r_1)-f_{2(I_{23})}(r_{1e}))\\
&+(f_{2(I_{24})}(r_1)-f_{2(I_{24})}(r_{1e})))
\end{split}
\end{equation*}

\subsubsection{\texorpdfstring{$0<r_i\leq r_{1e}$}{0<ri <= r1e}}
\begin{equation*}
\begin{split}
\mathcal{M}=&(1/(2\pi)) \times ((f_{1(I_{1})}(r_i)-f_{1(I_{1})}(0))
-f_{1n(I_{1})}\\
&+(f_{2(I_{1})}(r_{1e})-f_{2(I_{1})}(r_i))-f_{2n(I_{1})}\\
&+(f_{2(I_{21})}(r_1)-f_{2(I_{21})}(r_{1e}))-f_{2n(I_{21})}\\
&+(f_{2(I_{22})}(r_1)-f_{2(I_{22})}(r_{1e}))-f_{2n(I_{22})} \\
&+(f_{2(I_{23})}(r_1)-f_{2(I_{23})}(r_{1e}))-f_{2n(I_{23})}\\
&+(f_{2(I_{24})}(r_1)-f_{2(I_{24})}(r_{1e}))-f_{2n(I_{24})})            \end{split}
\end{equation*}

\subsubsection{\texorpdfstring{$r_{1e}<r_i<r_1$}{r1e<ri<r1}}
\begin{equation*}
    \begin{split}
\mathcal{M}=&(1/(2\pi)) \times ((f_{1(I_{21})}(r_i)-f_{1(I_{21})}(r_{1e}))-f_{1n(I_{21})} \\
&+(f_{1(I_{22})}(r_i)-f_{1(I_{22})}(r_{1e}))-f_{1n(I_{22})} \\
&+(f_{1(I_{23})}(r_i)-f_{1(I_{23})}(r_{1e}))-f_{1n(I_{23})} \\
&+(f_{1(I_{24})}(r_i)-f_{1(I_{24})}(r_{1e}))-f_{1n(I_{24})} \\
&+(f_{2(I_{21})}(r_1)-f_{2(I_{21})}(r_i))-f_{2n(I_{21})} \\
&+(f_{2(I_{22})}(r_1)-f_{2(I_{22})}(r_i))-f_{2n(I_{22})} \\
&+(f_{2(I_{23})}(r_1)-f_{2(I_{23})}(r_i))-f_{2n(I_{23})} \\
&+(f_{2(I_{24})}(r_1)-f_{2(I_{24})}(r_i))-f_{2n(I_{24})}\\
&+(f_{1(I_{1})}(r_{1e})-f_{1(I_{1})}(0))-f_{1n(I_{1})})        
    \end{split}
\end{equation*}

\subsubsection{\texorpdfstring{$r_i\geq r_1$}{ri >= r1}}
\begin{equation*}
    \begin{split}
\mathcal{M}=&(1/(2\pi))\times 
((f_{1(I_{1})}(r_{1e})-f_{1(I_{1})}(0))-f_{1n(I_{1})} \\
&+(f_{1(I_{21})}(r_1)-f_{1(I_{21})}(r_{1e}))-f_{1n(I_{21})} \\
&+(f_{1(I_{22})}(r_1)-f_{1(I_{22})}(r_{1e}))-f_{1n(I_{22})} \\
&+(f_{1(I_{23})}(r_1)-f_{1(I_{23})}(r_{1e}))-f_{1n(I_{23})} \\
&+(f_{1(I_{24})}(r_1)-f_{1(I_{24})}(r_{1e}))-f_{1n(I_{24})})
    \end{split}
\end{equation*}

\subsection{\texorpdfstring{Sub-functions of $f_1$}{Sub-functions of f1}}
\begin{equation*}
\begin{split}
f_{1(I_1)}&=-(\pi r^2\log(r_i)) \\
%\label{eq:wideeq}
%\end{equation}
%
%\begin{equation}
f_{1(I_{21})}&=(r(-(\kappa_{x_2}) - \kappa_{y_2} + r\alpha_2 - r\beta_2)\log(r_i))/2 \\
%\label{eq:wideeq}
%\end{equation}
%
%\begin{equation}
f_{1(I_{22})}&=(r(\kappa_{x_1} + \kappa_{y_2} - r\alpha_1 +        r\beta_2)\log(r_i))/2 \\
%\label{eq:wideeq}
%\end{equation}
%
%\begin{equation}
f_{1(I_{23})}&=(r(-(\kappa_{x_1}) - \kappa_{y_1} + r\alpha_1 -        r\beta_1)\log(r_i))/2 \\
%\label{eq:wideeq}
%\end{equation}
%
%\begin{equation}
f_{1(I_{24})}&=(r(\kappa_{x_2} + \kappa_{y_1} - r\alpha_2 +        r\beta_1)\log(r_i))/2 
%\label{eq:wideeq}
\end{split}
\end{equation*}

\subsection{\texorpdfstring{Sub-functions of $f_2$}{Sub-functions of f2}}
\begin{equation*}
f_{2(I_{1})}=2\pi(r^2/4 + (r^2\log(r^{-1}))/2)
%\label{eq:wideeq}
\end{equation*}

\begin{equation*}
\begin{split}
f_{2(I_{21})}=&(3r\epsilon_{x_2})/4 + (3r\epsilon_{y_2})/4 - (r^2\alpha_2)/4 + (x_2^2\gamma_2)/2 \\
&+(r^2\beta_2)/4 + (y_2^2\beta_2)/2
+ (r(\epsilon_{x_2} - r\alpha_2) \log(r^{-1}))/2 \\
&+ (r(\epsilon_{y_2} + r\beta_2) \log(r^{-1}))/2
\end{split}
%\label{eq:wideeq}
\end{equation*}

\begin{equation*}
\begin{split}
f_{2(I_{22})}=&(-3r\epsilon_{x_1})/4 - (3r\epsilon_{y_2})/4 + (r^2\alpha_1)/4 - (x_1^2\gamma_1)/2 \\
&- (r^2\beta_2)/4 - (y_2^2\beta_2)/2 
- (r(\epsilon_{x_1} - r\alpha_1) \log(r^{-1}))/2 \\
&- (r(\epsilon_{y_2} + r\beta_2) \log(r^{-1}))/2
\end{split}
%\label{eq:wideeq}
\end{equation*}

\begin{equation*}
\begin{split}
f_{2(I_{23})}=&(3r\epsilon_{x_1})/4 + (3r\epsilon_{y_1})/4 -      (r^2\alpha_1)/4 +  (x_1^2\gamma_1)/2 \\
&+ (r^2\beta_1)/4 +  (y_1^2\beta_1)/2 
+ (r(\epsilon_{x_1}- r\alpha_1)  \log(r^{-1}))/2 \\
&+ (r(\epsilon_{y_1} + r\beta_1) \log(r^{-1}))/2
\end{split}
%\label{eq:wideeq}
\end{equation*}

\begin{equation*}
\begin{split}
f_{2(I_{24})}=&(-3r\epsilon_{x_2})/4 - (3r\epsilon_{y_1})/4 
+ (r^2\alpha_2)/4 - (x_2^2\gamma_2)/2 \\
&- (r^2\beta_1)/4 - (y_1^2\beta_1)/2 
- (r(\epsilon_{x_2} - r\alpha_2)   \log(r^{-1}))/2 \\
&- (r(\epsilon_{y_1} + r\beta_1) \log(r^{-1}))/2
\end{split}
%\label{eq:wideeq}
\end{equation*}
%%%%%%%%%%%%%

\section{Coefficient calculation in \textit{Case 2}}\label{a:Case2}
Here are the formulas for calculating the coefficients $\mathcal{M}$ for actuators that can be categorized in \textit{Case 2}:

\subsubsection{\texorpdfstring{$r_i \leq r_1$}{ri <= r1}}
\begin{equation*}
    \begin{split}
\mathcal{M}=&(1/(2\pi))(((f_{2(I_{1})}(r_2)-f_{2(I_{1})}(r_1))-f_{2n(I_1)})\\
&+((f_{2(I_{2})}(r_4)-f_{2(I_{2})}(r_2))-f_{2n(I_2)}))
    \end{split}
\end{equation*}

\subsubsection{\texorpdfstring{$r_1<r_i\leq r_{2}$}{r1 < ri <= r2}}
\begin{equation*}
    \begin{split}
\mathcal{M}=&(1/(2\pi))(((f_{1(I_{1})}(r_i)-f_{1(I_{1})}(r_1))-f_{1n(I_1)})\\
&+((f_{2(I_{1})}(r_2)-f_{2(I_{1})}(r_i))-f_{2n(I_1)}) \\
&+((f_{2(I_{2})}(r_4)-f_{2(I_{2})}(r_2))-f_{2n(I_2)}))        
    \end{split}
\end{equation*}

\subsubsection{\texorpdfstring{$r_{2}<r_i<r_4$}{r2 < ri < r4}}
\begin{equation*}
    \begin{split}
\mathcal{M}=&(1/(2\pi))(((f_{1(I_{2})}(r_i)-f_{1(I_{2})}(r_2))-f_{1n(I_2)})\\
&+((f_{2(I_{2})}(r_4)-f_{2(I_{2})}(r_i))-f_{2n(I_2)})\\
&+((f_{1(I_{1})}(r_2)-f_{1(I_{1})}(r_1))-f_{1n(I_1)}))
    \end{split}
\end{equation*}

\subsubsection{\texorpdfstring{$r_i\geq r_4$}{ri >= r4}}
\begin{equation*}
    \begin{split}
\mathcal{M}=&(1/(2\pi))(((f_{1(I_{1})}(r_2)-f_{1(I_{1})}(r_1))-f_{1n(I_1)})\\
&+((f_{1(I_{2})}(r_4)-f_{1(I_{2})}(r_2))-f_{1n(I_2)}))         \end{split}
\end{equation*}

\subsection{\texorpdfstring{Sub-functions of $f_1$}{Sub-functions of f1}}
\begin{equation*}
\begin{split}
f_{1(I_1)}&=(r(\kappa_{x_1} + \kappa_{y_1} - r\alpha_1 +        r\beta_1)\log(r_i))/2 \\
%\end{split}
%\label{eq:wideeq}
%\end{equation}
%
%\begin{equation}
%\begin{split}
f_{1(I_2)}&=(r(-(\kappa_{x_2}) - \kappa_{y_2} + r\alpha_2 -        r\beta_2)\log(r_i))/2 
\end{split}
%\label{eq:wideeq}
\end{equation*}

\subsection{\texorpdfstring{Sub-functions of $f_2$}{Sub-functions of f2}}
\begin{equation*}
\begin{split}
f_{2(I_1)}=&(-3r\epsilon_{x_1})/4 - (3r\epsilon_{y_1})/4 
+  (r^2\alpha_1)/4 - (x_1^2\gamma_1)/2 \\
&- (r^2\beta_1)/4 -  (y_1^2\beta_1)/2 - (r(\epsilon_{x_1} - r\alpha_1) \log(r^{-1}))/2 \\
&- (r(\epsilon_{y_1} + r\beta_1)   \log(r^{-1}))/2
\end{split}
%\label{eq:wideeq}
\end{equation*}

\begin{equation*}
\begin{split}
f_{2(I_2)}=&(3r\epsilon_{x_2})/4 + (3r\epsilon_{y_2})/4 
- (r^2\alpha_2)/4 + (x_2^2\gamma_2)/2\\
&+ (r^2\beta_2)/4 + (y_2^2\beta_2)/2 + (r(\epsilon_{x_2} - r\alpha_2) \log(r^{-1}))/2 \\
&+ (r(\epsilon_{y_2} + r\beta_2)   \log(r^{-1}))/2
\end{split}
%\label{eq:wideeq}
\end{equation*}
%%%%%%%%%%%%%

\section{Coefficient calculation in \textit{Case 3}}\label{a:Case3}
Here are the formulas for calculating the coefficients $\mathcal{M}$ for actuators that can be categorized in \textit{Case 3}:

\subsubsection{\texorpdfstring{$r_i \leq r_{1e}$}{ri <= r1e}}
\begin{equation*}
    \begin{split}
\mathcal{M}=&(1/(2\pi))((f_{2(I_{11})}(r_1)-f_{2(I_{11})}(r_{1e}))-f_{2n(I_{11})} \\
&+(f_{2(I_{12})}(r_1)-f_{2(I_{12})}(r_{1e}))-f_{2n(I_{12})} \\
&+(f_{2(I_{2})}(r_{4e})-f_{2(I_{2})}(r_1))-f_{2n(I_{2})} \\
&+(f_{2(I_{31})}(r_4)-f_{2(I_{31})}(r_{4e}))-f_{2n(I_{31})} \\
&+(f_{2(I_{32})}(r_4)-f_{2(I_{32})}(r_{4e}))-f_{2n(I_{32})})
    \end{split}
\end{equation*} 

\subsubsection{\texorpdfstring{$r_{1e}<r_i\leq r_{1}$}{r1e < ri <= r1}}
\begin{equation*}
    \begin{split}
\mathcal{M}=&(1/(2\pi))((f_{1(I_{11})}(r_i)-f_{1(I_{11})}(r_{1e}))-f_{1n(I_{11})} \\
&+(f_{1(I_{12})}(r_i)-f_{1(I_{12})}(r_{1e}))-f_{1n(I_{12})} \\
&+(f_{2(I_{11})}(r_1)-f_{2(I_{11})}(r_i))-f_{2n(I_{11})} \\
&+(f_{2(I_{12})}(r_1)-f_{2(I_{12})}(r_i))-f_{2n(I_{12})}\\
&+(f_{2(I_{2})}(r_{4e})-f_{2(I_{2})}(r_1))-f_{2n(I_{2})} \\
&+(f_{2(I_{31})}(r_4)-f_{2(I_{31})}(r_{4e}))-f_{2n(I_{31})} \\
&+(f_{2(I_{32})}(r_4)-f_{2(I_{32})}(r_{4e}))-f_{2n(I_{32})})    
    \end{split}
\end{equation*}

\subsubsection{\texorpdfstring{$r_{1}<r_i \leq r_{4e}$}{r1 < ri <= r4e}}
\begin{equation*}
    \begin{split}
\mathcal{M}=&(1/(2\pi))((f_{1(I_{2})}(r_i)-f_{1(I_{2})}(r_1))-f_{1n(I_{2})} \\
&+(f_{2(I_{2})}(r_{4e})-f_{2(I_{2})}(r_i))-f_{2n(I_{2})}\\
&+(f_{1(I_{11})}(r_1)-f_{1(I_{11})}(r_{1e}))-f_{1n(I_{11})} \\
&+(f_{1(I_{12})}(r_1)-f_{1(I_{12})}(r_{1e}))-f_{1n(I_{12})}\\
&+(f_{2(I_{31})}(r_4)-f_{2(I_{31})}(r_{4e}))-f_{2n(I_{31})}\\
&+(f_{2(I_{32})}(r_4)-f_{2(I_{32})}(r_{4e}))-f_{2n(I_{32})})
    \end{split}
\end{equation*}

\subsubsection{\texorpdfstring{$r_{4e}<r_i < r_{4}$}{r4e < ri < r4}}
\begin{equation*}
    \begin{split}
\mathcal{M}=&(1/(2\pi))((f_{1(I_{31})}(r_i)-f_{1(I_{31})}(r_{4e}))-f_{1n(I_{31})}\\
&+(f_{1(I_{32})}(r_i)-f_{1(I_{32})}(r_{4e}))-f_{1n(I_{32})} \\
&+(f_{2(I_{31})}(r_4)-f_{2(I_{31})}(r_i))-f_{2n(I_{31})}\\
&+(f_{2(I_{32})}(r_4)-f_{2(I_{32})}(r_i))-f_{2n(I_{32})}\\
&+(f_{1(I_{11})}(r_1)-f_{1(I_{11})}(r_{1e}))-f_{1n(I_{11})} \\
&+(f_{1(I_{12})}(r_1)-f_{1(I_{12})}(r_{1e}))-f_{1n(I_{12})}\\
&+(f_{1(I_{2})}(r_{4e})-f_{1(I_{2})}(r_1))-f_{1n(I_{2})})        
    \end{split}
\end{equation*}

\subsubsection{\texorpdfstring{$r_i \geq r_4$}{ri >= r4}}
\begin{equation*}
    \begin{split}
\mathcal{M}=&(1/(2\pi))((f_{1(I_{11})}(r_1)-f_{1(I_{11})}(r_{1e}))-f_{1n(I_{11})}\\
&+(f_{1(I_{12})}(r_1)-f_{1(I_{12})}(r_{1e}))-f_{1n(I_{12})} \\
&+(f_{1(I_{2})}(r_{4e})-f_{1(I_{2})}(r_1))-f_{1n(I_{2})}\\
&+(f_{1(I_{31})}(r_4)-f_{1(I_{31})}(r_{4e}))-f_{1n(I_{31})} \\
&+(f_{1(I_{32})}(r_4)-f_{1(I_{32})}(r_{4e}))-f_{1n(I_{32})})      
    \end{split}
\end{equation*}

\subsection{\texorpdfstring{Sub-functions of $f_1$}{Sub-functions of f1}}
\begin{equation*}
\begin{split}
f_{1(I_{11})}&=-((-(r\kappa_{x_1}) + r^2\alpha_1)\log(r_i))/2 \\
%\end{split}
%\label{eq:wideeq}
%\end{equation}
%
%\begin{equation}
%\begin{split}
f_{1(I_{12})}&=-((-(r\kappa_{x_1}) + r^2\alpha_1)\log(r_i))/2 \\
%\end{split}
%\label{eq:wideeq}
%\end{equation}
%
%\begin{equation}
%\begin{split}
f_{1(I_{2})}&=(r(\kappa_{y_1} - \kappa_{y_2} + r\beta_1 -        r\beta_2)\log(r_i))/2 \\
%\end{split}
%\label{eq:wideeq}
%\end{equation}
%
%\begin{equation}
%\begin{split}
f_{1(I_{31})}&=(r(-(\kappa_{x_2}) - \kappa_{y_2} + r\alpha_2 - r\beta_2)\log(r_i))/2 \\
%\end{split}
%\label{eq:wideeq}
%\end{equation}
%
%\begin{equation}
%\begin{split}
f_{1(I_{32})}&=(r(\kappa_{x_2} + \kappa_{y_1} - r\alpha_2 +        r\beta_1)\log(r_i))/2 
\end{split}
%\label{eq:wideeq}
\end{equation*}

\subsection{\texorpdfstring{Sub-functions of $f_2$}{Sub-functions of f2}}
\begin{equation*}
\begin{split}
f_{2(I_{11})}=&(-2x_1^2\gamma_1 + r^2\alpha_1(1 + 2\log(r^{-1})) \\
&- r\kappa_{x_1}(3 + 2\log(r^{-1})))/4
\end{split}
%\label{eq:wideeq}
\end{equation*}

\begin{equation*}
\begin{split}
f_{2(I_{12})}=&(-2x_1^2\gamma_1 + r^2\alpha_1(1 + 2\log(r^{-1}))\\
&- r\kappa_{x_1}(3 + 2\log(r^{-1})))/4
\end{split}
%\label{eq:wideeq}
\end{equation*}

\begin{equation*}
\begin{split}
f_{2(I_{2})}=&(-(r(\kappa_{y_1} - \kappa_{y_2})(3 + 2\log(r^{-1})))
- \beta_1(r^2 + 2y_1^2 \\
&+ 2r^2\log(r^{-1}))  
+\beta_2(r^2 + 2y_2^2 + 2r^2\log(r^{-1})))/4
\end{split}
%\label{eq:wideeq}
\end{equation*}

\begin{equation*}
\begin{split}
f_{2(I_{31})}=&(3r\epsilon_{x_2})/4 + (3r\epsilon_{y_2})/4
- (r^2\alpha_2)/4 + (x_2^2\gamma_2)/2 \\
&+ (r^2\beta_2)/4 + (y_2^2\beta_2)/2 
+ (r(\epsilon_{x_2} - r\alpha_2) \log(r^{-1}))/2\\
&+ (r(\epsilon_{y_2} + r\beta_2)  \log(r^{-1}))/2
\end{split}
%\label{eq:wideeq}
\end{equation*}

\begin{equation*}
\begin{split}
f_{2(I_{32})}=&(-3r\epsilon_{x_2})/4 - (3r\epsilon_{y_1})/4 
+ (r^2\alpha_2)/4 - (x_2^2\gamma_2)/2 \\
&- (r^2\beta_1)/4 -  (y_1^2\beta_1)/2 
- (r(\epsilon_{x_2} - r\alpha_2) \log(r^{-1}))/2 \\
&- (r(\epsilon_{y_1} + r\beta_1) \log(r^{-1}))/2
\end{split}
%\label{eq:wideeq}
\end{equation*}
%%%%%%%%%%%%%

\section{Coefficient calculation in \textit{Case 4}}\label{a:Case4}
Here are the formulas for calculating the coefficients $\mathcal{M}$ for actuators that can be categorized in \textit{Case 4}:

\subsubsection{\texorpdfstring{$r_i \leq r_{1}$}{ri <= r1}}
\begin{equation*}
    \begin{split}
\mathcal{M}=&(1/(2\pi))(((f_{2(I_{1})}(r_3)-f_{2(I_{1})}(r_1))-f_{2n(I_{1})})\\
&+((f_{2(I_{2})}(r_2)-f_{2(I_{2})}(r_3))-f_{2n(I_{2})})\\
&+((f_{2(I_{3})}(r_4)-f_{2(I_{3})}(r_2))-f_{2n(I_{3})}))        
    \end{split}
\end{equation*}

\subsubsection{\texorpdfstring{$r_{1}<r_i\leq r_{3}$}{r1 < ri <= r3}}
\begin{equation*}
    \begin{split}
\mathcal{M}=&(1/(2\pi))(((f_{1(I_{1})}(r_i)-f_{1(I_{1})}(r_1))-f_{1n(I_{1})})\\
&+((f_{2(I_{1})}(r_3)-f_{2(I_{1})}(r_i))-f_{2n(I_{1})}) \\
&+((f_{2(I_{2})}(r_2)-f_{2(I_{2})}(r_3))-f_{2n(I_{2})})\\
&+((f_{2(I_{3})}(r_4)-f_{2(I_{3})}(r_2))-f_{2n(I_{3})}))        
    \end{split}
\end{equation*}

\subsubsection{\texorpdfstring{$r_{3}<r_i \leq r_{2}$}{r3 < ri <= r2}}
\begin{equation*}
    \begin{split}
\mathcal{M}=&(1/(2\pi))(((f_{1(I_{2})}(r_i)-f_{1(I_{2})}(r_3))-f_{1n(I_{2})})\\
&+((f_{2(I_{2})}(r_2)-f_{2(I_{2})}(r_i))-f_{2n(I_{2})}) \\
&+((f_{1(I_{1})}(r_3)-f_{1(I_{1})}(r_1))-f_{1n(I_{1})})\\
&+((f_{2(I_{3})}(r_4)-f_{2(I_{3})}(r_2))-f_{2n(I_{3})}))        
    \end{split}
\end{equation*}

\subsubsection{\texorpdfstring{$r_{2}<r_i < r_{4} $}{r2 < ri < r4}}
\begin{equation*}
    \begin{split}
\mathcal{M}=&(1/(2\pi))(((f_{1(I_{3})}(r_i)-f_{1(I_{3})}(r_2))-f_{1n(I_{3})})\\
&+((f_{2(I_{3})}(r_4)-f_{2(I_{3})}(r_i))-f_{2n(I_{3})}) \\
&+((f_{1(I_{1})}(r_3)-f_{1(I_{1})}(r_1))-f_{1n(I_{1})})\\
&+((f_{1(I_{2})}(r_2)-f_{1(I_{2})}(r_3))-f_{1n(I_{2})}))        
    \end{split}
\end{equation*}

\subsubsection{\texorpdfstring{$r_i \geq r_4$}{ri >= r4}}
\begin{equation*}
    \begin{split}
\mathcal{M}=&(1/(2\pi))(((f_{1(I_{1})}(r_3)-f_{1(I_{1})}(r_1))-f_{1n(I_{1})})\\
&+((f_{1(I_{2})}(r_2)-f_{1(I_{2})}(r_3))-f_{1n(I_{2})}) \\
&+((f_{1(I_{3})}(r_4)-f_{1(I_{3})}(r_2))-f_{1n(I_{3})}))
    \end{split}
\end{equation*}

\subsection{\texorpdfstring{Sub-functions of $f_1$}{Sub-functions of f1}}
\begin{equation*}
\begin{split}
f_{1(I_{1})}&=(r(\kappa_{x_1} + \kappa_{y_1} - r\alpha_1 +        r\beta_1)\log(r_i))/2 \\
%\end{split}
%\label{eq:wideeq}
%\end{equation}
%
%\begin{equation}
%\begin{split}
f_{1(I_{2})}&=(r(\kappa_{x_1} - \kappa_{x_2} - r\alpha_1 +        r\alpha_2)\log(r_i))/2 \\
%\end{split}
%\label{eq:wideeq}
%\end{equation}
%
%\begin{equation}
%\begin{split}
f_{1(I_{3})}&=(r(-(\kappa_{x_2}) - \kappa_{y_2} + r\alpha_2 -        r\beta_2)\log(r_i))/2
\end{split}
%\label{eq:wideeq}
\end{equation*}

\subsection{\texorpdfstring{Sub-functions of $f_2$}{Sub-functions of f2}}
\begin{equation*}
\begin{split}
f_{2(I_{1})}=&(-3r\epsilon_{x_1})/4 - (3r\epsilon_{y_1})/4 
+ (r^2\alpha_1)/4 - (x_1^2\gamma_1)/2 \\
&- (r^2\beta_1)/4 -  (y_1^2\beta_1)/2 
- (r(\epsilon_{x_1} - r\alpha_1)       \log(r^{-1}))/2 \\
&- (r(\epsilon_{y_1} + r\beta_1)       \log(r^{-1}))/2
\end{split}
%\label{eq:wideeq}
\end{equation*}

\begin{equation*}
\begin{split}
f_{2(I_{2})}=&(-3r\epsilon_{x_1})/4 + (3r\epsilon_{x_2})/4 
+ (r^2\alpha_1)/4  - (r^2\alpha_2)/4  \\
&- (x_1^2\gamma_1)/2 + (x_2^2\gamma_2)/2 
- (r(\epsilon_{x_1} - r\alpha_1)       \log(r^{-1}))/2 \\
&+ (r(\epsilon_{x_2} - r\alpha_2)       \log(r^{-1}))/2
\end{split}
%\label{eq:wideeq}
\end{equation*}

\begin{equation*}
\begin{split}
f_{2(I_{3})}=&(3r\epsilon_{x_2})/4 + (3r\epsilon_{y_2})/4 
- (r^2\alpha_2)/4 + (x_2^2\gamma_2)/2 \\
&+ (r^2\beta_2)/4 + (y_2^2\beta_2)/2 
+ (r(\epsilon_{x_2} - r\alpha_2)       \log(r^{-1}))/2 \\
&+ (r(\epsilon_{y_2} + r\beta_2)       \log(r^{-1}))/2
\end{split}
%\label{eq:wideeq}
\end{equation*}

%%%%%%%%%%%%%
\section{Coefficient calculation in \textit{Case 5}}\label{a:Case5}
Here are the formulas for calculating the coefficients $\mathcal{M}$ for actuators that can be categorized in \textit{Case 5}:

\subsubsection{\texorpdfstring{$r_i \leq r_{1}$}{ri <= r1}}
\begin{equation*}
    \begin{split}
\mathcal{M}=&(1/(2\pi))(((f_{2(I_{1})}(r_2)-f_{2(I_{1})}(r_1))-f_{2n(I_1)})\\
&+((f_{2(I_{2})}(r_3)-f_{2(I_{2})}(r_2))-f_{2n(I_2)}) \\
&+((f_{2(I_{3})}(r_4)-f_{2(I_{3})}(r_3))-f_{2n(I_3)}))
    \end{split}
\end{equation*}

\subsubsection{\texorpdfstring{$r_{1}<r_i\leq r_{2}$}{r1 < ri <= r2}}
\begin{equation*}
    \begin{split}
\mathcal{M}=&(1/(2\pi))(((f_{1(I_{1})}(r_i)-f_{1(I_{1})}(r_1))-f_{1n(I_1)})\\
&+((f_{2(I_{1})}(r_2)-f_{2(I_{1})}(r_i))-f_{2n(I_1)}) \\
&+((f_{2(I_{2})}(r_3)-f_{2(I_{2})}(r_2))-f_{2n(I_2)}) \\
&+((f_{2(I_{3})}(r_4)-f_{2(I_{3})}(r_3))-f_{2n(I_3)}))
    \end{split}
\end{equation*}

\subsubsection{\texorpdfstring{$r_{2}<r_i \leq r_{3}$}{r2 < ri <= r3}}
\begin{equation*}
    \begin{split}
\mathcal{M}=&(1/(2\pi))(((f_{1(I_{2})}(r_i)-f_{1(I_{2})}(r_2))-f_{1n(I_2)}) \\
&+((f_{2(I_{2})}(r_3)-f_{2(I_{2})}(r_i))-f_{2n(I_2)}) \\
&+((f_{1(I_{1})}(r_2)-f_{1(I_{1})}(r_1))-f_{1n(I_1)}) \\
&+((f_{2(I_{3})}(r_4)-f_{2(I_{3})}(r_3))-f_{2n(I_3)}))
    \end{split}
\end{equation*}

\subsubsection{\texorpdfstring{$r_{3}<r_i < r_{4} $}{r3 < ri < r4}}
\begin{equation*}
    \begin{split}
\mathcal{M}=&(1/(2\pi))(((f_{1(I_{3})}(r_i)-f_{1(I_{3})}(r_3))-f_{1n(I_3)}) \\
&+((f_{2(I_{3})}(r_4)-f_{2(I_{3})}(r_i))-f_{2n(I_3)}) \\
&+((f_{1(I_{1})}(r_2)-f_{1(I_{1})}(r_1))-f_{1n(I_1)}) \\
&+((f_{1(I_{2})}(r_3)-f_{1(I_{2})}(r_2))-f_{1n(I_2)}))
    \end{split}
\end{equation*}

\subsubsection{\texorpdfstring{$r_i \geq r_4$}{ri >= r4}}
\begin{equation*}
    \begin{split}
\mathcal{M}=&(1/(2\pi))(((f_{1(I_{1})}(r_2)-f_{1(I_{1})}(r_1))-f_{1n(I_1)}) \\
&+((f_{1(I_{2})}(r_3)-f_{1(I_{2})}(r_2))-f_{1n(I_2)})\\
&+((f_{1(I_{3})}(r_4)-f_{1(I_{3})}(r_3))-f_{1n(I_3)}))
    \end{split}
\end{equation*}

\subsection{\texorpdfstring{Sub-functions of $f_1$}{Sub-functions of f1}}
\begin{equation*}
\begin{split}
f_{1(I_{1})}&=(r(\kappa_{x_1} + \kappa_{y_1} - r\alpha_1 +        r\beta_1)\log(r_i))/2 \\
%\end{split}
%\label{eq:wideeq}
%\end{equation}
%
%\begin{equation}
%\begin{split}
f_{1(I_{2})}&=(r(\kappa_{y_1} - \kappa_{y_2} + r\beta_1 -        r\beta_2)\log(r_i))/2 \\
%\end{split}
%\label{eq:wideeq}
%\end{equation}
%
%\begin{equation}
%\begin{split}
f_{1(I_{3})}&=(r(-(\kappa_{x_2}) - \kappa_{y_2} + r\alpha_2 -        r\beta_2)\log(r_i))/2 
\end{split}
%\label{eq:wideeq}
\end{equation*}

\subsection{\texorpdfstring{Sub-functions of $f_2$}{Sub-functions of f2}}
\begin{equation*}
\begin{split}
f_{2(I_{1})}=&(-3r\epsilon_{x_1})/4 - (3r\epsilon_{y_1})/4 
+ (r^2\alpha_1)/4 - (x_1^2\gamma_1)/2 \\
&- (r^2\beta_1)/4 - (y_1^2\beta_1)/2 
- (r(\epsilon_{x_1} - r\alpha_1)       \log(r^{-1}))/2 \\
&- (r(\epsilon_{y_1} + r\beta_1)       \log(r^{-1}))/2
\end{split}
%\label{eq:wideeq}
\end{equation*}

\begin{equation*}
\begin{split}
f_{2(I_{2})}=&(-(r(\kappa_{y_1} - \kappa_{y_2})(3 + 2\log(r^{-1})))
- \beta_1(r^2 + 2y_1^2 \\
&+ 2r^2\log(r^{-1})) 
+ \beta_2(r^2 + 2y_2^2 + 2r^2\log(r^{-1})))/4
\end{split}
%\label{eq:wideeq}
\end{equation*}

\begin{equation*}
\begin{split}
f_{2(I_{3})}=&(3r\epsilon_{x_2})/4 + (3r\epsilon_{y_2})/4 
- (r^2\alpha_2)/4 + (x_2^2\gamma_2)/2\\
&+ (r^2\beta_2)/4 + (y_2^2\beta_2)/2 
+ (r(\epsilon_{x_2} - r\alpha_2)       \log(r^{-1})/2 \\
&+ (r(\epsilon_{y_2} + r\beta_2)       \log(r^{-1}))/2
\end{split}
%\label{eq:wideeq}
\end{equation*}

%APPENDIX
%%%%%%%%%%%%%%%%%%%%%%%%%%%%%%%%%%%%%%%%%%%%%%%%%%%%%%%%%%%%%%%%%%%%%
\section*{Funding}
This work was co-funded through a Marie Skłodowska-Curie COFUND (DSSC 754315).

\section*{Acknowledgments}
The authors thank Dr. M. Acuautla and Prof. B. Noheda of the University of Groningen, and Dr. S.N.R. Kazmi, M. Eggens and H. Smit of the Netherlands Institute for Space Research for their valuable input on the Hysteretic Deformable Mirror project. In addition, we would like to thank the Center for Information Technology of the University of Groningen for their support and for providing access to the Peregrine high performance computing cluster.

\section*{Disclosure}
The authors declare no conflicts of interest.
%%%%%%%%%%%%%%%%%%%%%%%%%%%%%%%%%%%%%%%%%%%%%%%%%%%%%%%%%%%%%%%%%%%%%

% Bibliography
\bibliography{sample}

% Full bibliography added automatically for Optics Letters submissions; the following line will simply be ignored if submitting to other journals.
% Note that this extra page will not count against page length
\bibliographyfullrefs{sample}

%Manual citation list
%\begin{thebibliography}{1}
%\bibitem{Zhang:14}
%Y.~Zhang, S.~Qiao, L.~Sun, Q.~W. Shi, W.~Huang, %L.~Li, and Z.~Yang,
 % \enquote{Photoinduced active terahertz metamaterials with nanostructured
  %vanadium dioxide film deposited by sol-gel method,} Opt. Express \textbf{22},
  %11070--11078 (2014).
%\end{thebibliography}

% Please include bios and photos of all authors for aop articles
%\ifthenelse{\equal{\journalref}{aop}}{%
%\section*{Author Biographies}
%\begingroup
%\setlength\intextsep{0pt}
%\begin{minipage}[t][6.3cm][t]{1.0\textwidth} % Adjust height [6.3cm] as required for separation of bio photos.
%  \begin{wrapfigure}{L}{0.25\textwidth}
%    \includegraphics[width=0.25\textwidth]{john_smith.eps}
%  \end{wrapfigure}
%  \noindent
%  {\bfseries John Smith} received his BSc (Mathematics) in 2000 from The University of Maryland. His research interests include lasers and optics.
%\end{minipage}
%\begin{minipage}{1.0\textwidth}
%  \begin{wrapfigure}{L}{0.25\textwidth}
%    \includegraphics[width=0.25\textwidth]{alice_smith.eps}
%  \end{wrapfigure}
%  \noindent
%  {\bfseries Alice Smith} also received her BSc (Mathematics) in 2000 from The University of Maryland. Her research interests also include lasers and optics.
%\end{minipage}
%\endgroup
%}{}

\end{document}